\documentclass[sigconf]{acmart}
\AtBeginDocument{%
  }

\setcopyright{acmlicensed}
\copyrightyear{2023}
\acmYear{2023}
\acmDOI{10.1145/3581783.3612084}

\acmConference[MM '23]{Proceedings of the 31st ACM International Conference on Multimedia}{October 29--November 3, 2023}{Ottawa, ON, Canada.}
\acmBooktitle{Proceedings of the 31st ACM International Conference on Multimedia (MM '23), October 29--November 3, 2023, Ottawa, ON, Canada}
\acmPrice{15.00}
\acmISBN{979-8-4007-0108-5/23/10}

\begin{document}


\title{U2Net: A General Framework with Spatial-Spectral-Integrated Double U-Net for Image Fusion}


\author{Siran Peng}
\authornote{Equal contribution.}
\affiliation{%
  \institution{School of Information and Communication Engineering, University of Electronic Science and Technology of China}
  \city{}
  \country{}
}
\email{Siran_Peng@163.com}

\author{Chenhao Guo}
\authornotemark[1]
\affiliation{%
	\institution{School of Information and Communication Engineering, University of Electronic Science and Technology of China}
	\city{}
	\country{}
}
\email{carlguo508@gmail.com}

\author{Xiao Wu}
\affiliation{%
  \institution{School of Mathematical Sciences, University of Electronic Science and Technology of China}
  \city{}
  \country{}
}
\email{wxwsx1997@gmail.com}

\author{Liang-Jian Deng}
\authornote{Corresponding author.}
\affiliation{%
	\institution{School of Mathematical Sciences, University of Electronic Science and Technology of China}
	\city{}
	\country{}
}
\email{liangjian.deng@uestc.edu.cn}

\renewcommand{\shortauthors}{Siran Peng, Chenhao Guo, Xiao Wu, \& Liang-Jian Deng}

\begin{abstract}

In image fusion tasks, images obtained from different sources exhibit distinct properties. Consequently, treating them uniformly with a single-branch network can lead to inadequate feature extraction. Additionally, numerous works have demonstrated that multi-scaled networks capture information more sufficiently than single-scaled models in pixel-level computer vision problems. Considering these factors, we propose U2Net, a spatial-spectral-integrated double U-shape network for image fusion. The U2Net utilizes a spatial U-Net and a spectral U-Net to extract spatial details and spectral characteristics, which allows for the discriminative and hierarchical learning of features from diverse images. In contrast to most previous works that merely employ concatenation to merge spatial and spectral information, this paper introduces a novel spatial-spectral integration structure called S2Block, which combines feature maps from different sources in a logical and effective way. We conduct a series of experiments on two image fusion tasks, including remote sensing pansharpening and hyperspectral image super-resolution (HISR). The U2Net outperforms representative state-of-the-art (SOTA) approaches in both quantitative and qualitative evaluations, demonstrating the superiority of our method. The code is available at \url{https://github.com/PSRben/U2Net}.

\end{abstract}




\maketitle

\begin{figure}[t]
	\begin{center}
		\begin{minipage}{1\linewidth}
			{\includegraphics[width=0.92\linewidth]{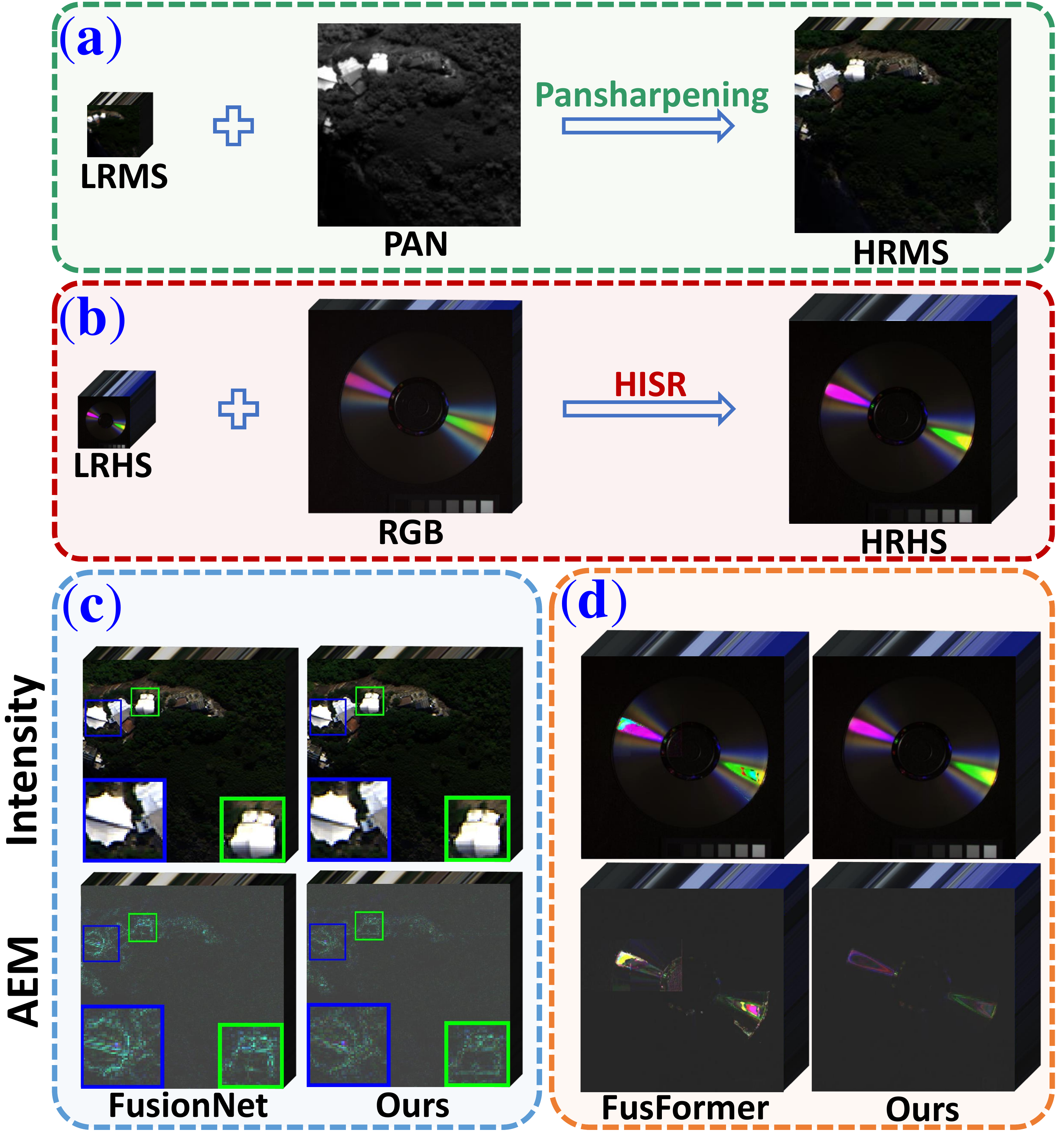}}
			\centering
		\end{minipage}
	\end{center}
	\caption{
		(a) The schematic diagram of pansharpening. (b) The schematic diagram of HISR. (c) The pansharpened images and their absolute error maps (AEMs) of FusionNet \cite{2020Detail} and U2Net. (d) The acquired HRHS images and corresponding AEMs of Fusformer \cite{9841513} and U2Net. It is obvious that our method yields the darker AEMs on both image fusion tasks, indicating its superiority over other competitors.\label{hp}}
\end{figure}

\section{Introduction}
Due to hardware limitations, sensors can only acquire high resolution images with sparse spectral information and low resolution images with copious spectral data. Image fusion aims to combine these two kinds of images to produce high resolution results with a wealth of spectral information. Over recent years, image fusion algorithms have been widely used in fields such as remote sensing \cite{vivone2014critical}, medical imaging \cite{james2014medical}, and computer vision
\cite{10.1145/3343031.3350928}, proving high application values. This work mainly investigates two image fusion tasks: remote sensing pansharpening and hyperspectral image super-resolution (HISR). As illustrated in Fig.~\ref{hp}, pansharpening involves merging a panchromatic (PAN) image with a low resolution multispectral (LRMS) image to create a high resolution multispectral (HRMS) outcome, while HISR aims at generating a high resolution hyperspectral (HRHS) result from an RGB image and a low resolution hyperspectral (LRHS) image.

The traditional pansharpening works can be roughly divided into three categories \cite{2020Detail}, \emph{i.e.}, the component substitution (CS) approaches, the multi-resolution analysis (MRA) methods, and the variational optimization-based (VO) techniques. The CS-based approaches \cite{2010A,2019Robust} project the LRMS image into a transformed domain, where the spatial information can be viewed as a component. By replacing this component with the PAN image, a desired HRMS result is generated. Although the CS-based approaches offer simple operation, low computational burden, and high spatial fidelity, they often suffer from significant spectral distortions.
The MRA-based methods \cite{6616569,vivone2018full} utilize a multi-resolution analysis framework to inject spatial details from the PAN image into the LRMS image, resulting in an HRMS output. While these methods maintain spectral characteristics effectively, they may encounter spatial distortion.
The VO-based techniques \cite{palsson2013new,6845349,10.1145/3503161.3547774,yan2022panchromatic} exploit different optimization algorithms to solve the pansharpening issue and generally outperform CS-based and MRA-based approaches. Nevertheless, these techniques have problems such as high computational burden and complex parameterization, which restrict their practical implementation. As for the HISR task, conventional approaches focus on exploring the inherent relationship between the RGB and LRHS images and mainly establish models based on optimization to obtain the HRHS image.

Over the past few years, deep learning (DL) has emerged as a popular solution for image fusion problems. Thanks to the exceptional feature learning capacity of neural networks, numerous DL-based methods have yielded impressive outcomes. The classic DL-based approaches \cite{8237455,2018Pansharpening,9841513,10.1145/3474085.3475571} apply concatenation to combine images from different sources. The cascaded output is then fed into a single-branch, single-scale network to generate a desired outcome. However, this strategy suffers from several significant defects. Firstly, the concatenation operation fails to consider the distinctions between two types of images, causing insufficient information integration. Secondly, the single-branch design leads to inefficient feature extraction as it treats spatial and spectral characteristics equally. Thirdly, some deep-level information may be ignored due to single-scale image processing.

To address the abovementioned concerns, we propose a spatial-spectral-integrated double U-shape network called U2Net for image fusion. The U2Net utilizes a spatial U-Net to capture spatial details from the PAN/RGB image and employs a spectral U-Net to extract spectral characteristics from the LRMS/LRHS image. This enables our method to learn diverse features in a discriminative and hierarchical manner. Besides, a novel structure named S2Block is introduced to integrate the two kinds of information. In the S2Block, we first generate two sets of square matrices, namely \emph{spatial self-correlation matrices} and \emph{spectral self-correlation matrices}, to better describe the spatial and spectral information. Subsequently, a series of operators are applied to combine the square matrices, spatial feature maps, and spectral feature maps, producing a high-quality fusion result. The contributions of this work are as follows:

\begin{itemize}
	\item A double U-shape network architecture consisting of a spatial U-Net and a spectral U-Net is created for image fusion tasks. This framework enables the effective learning of spatial details and spectral characteristics in a discriminative and hierarchical manner.
	
	\item A novel spatial-spectral integration structure called S2Block is designed to sufficiently merge feature maps from diverse images in a logical and comprehensive way. 
	
	\item The spatial U-Net and spectral U-Net are connected through S2Blocks, composing our U2Net. The proposed method is tested on different image fusion tasks and achieves SOTA performance in quantitative and qualitative assessments.
\end{itemize}

\begin{figure}[t]
	\begin{center}
		\begin{minipage}{1\linewidth}
			{\includegraphics[width=0.90\linewidth]{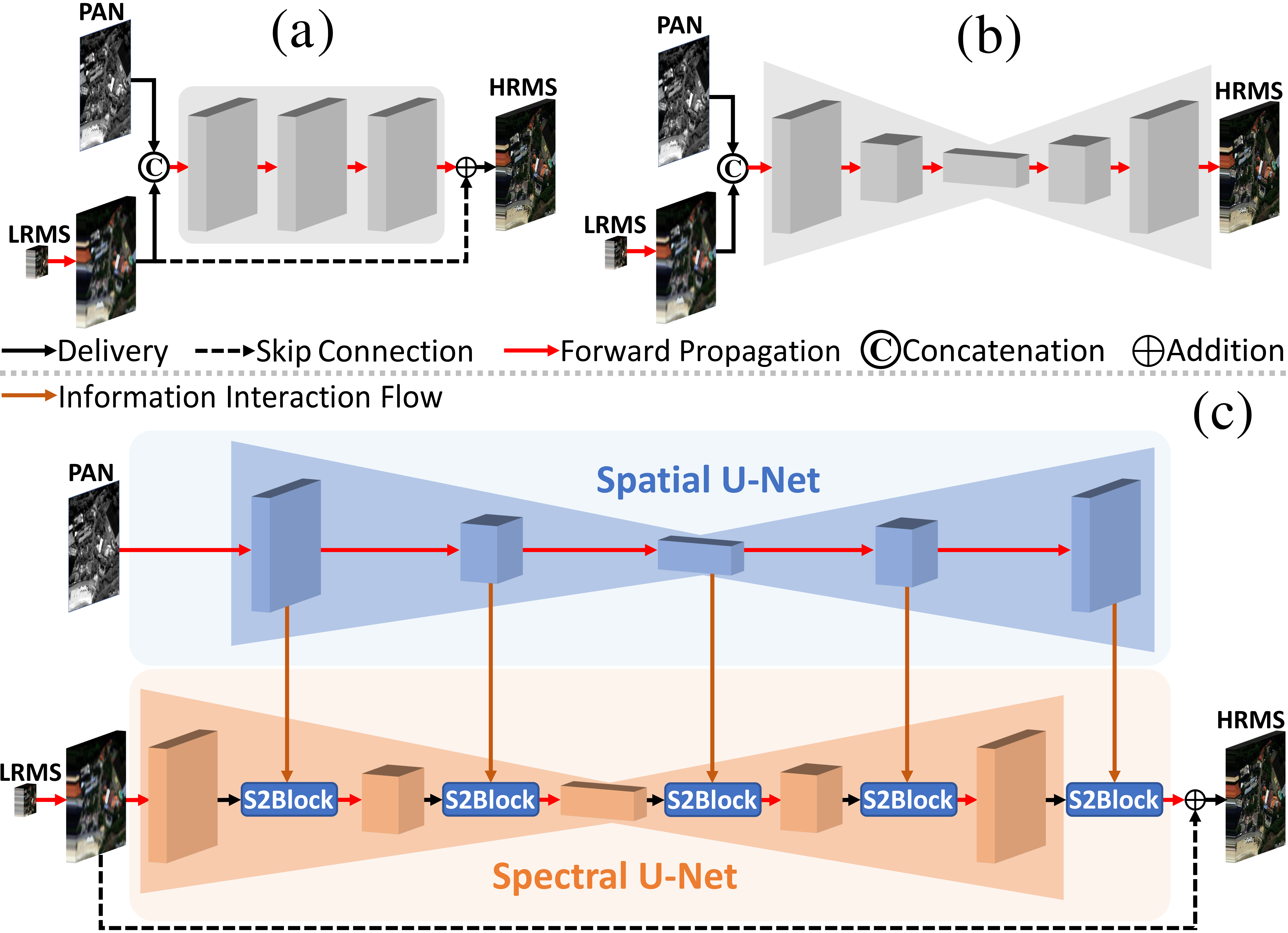}}
			\centering
		\end{minipage}
	\end{center}
	\caption{The structural comparison between existing DL-based image fusion works and U2Net (demonstrated with the pansharpening task). (a) The overall architecture of single-branch, single-scale methods, such as PanNet \cite{8237455} and Fusformer \cite{9841513}. These methods extract features at one specific scale, thus ignoring some deep-level information. (b) The overall architecture of single-branch, multi-scale approaches, including DCFNet \cite{Wu_2021_ICCV} and MUCNN \cite{10.1145/3474085.3475600}. (c) The overall structure of the double-branch, multi-scale U2Net.}
	\label{cmprr}
\end{figure}


\begin{figure*}[t]
	\begin{center}
		\begin{minipage}{1\linewidth}
			{\includegraphics[width=0.93\linewidth]{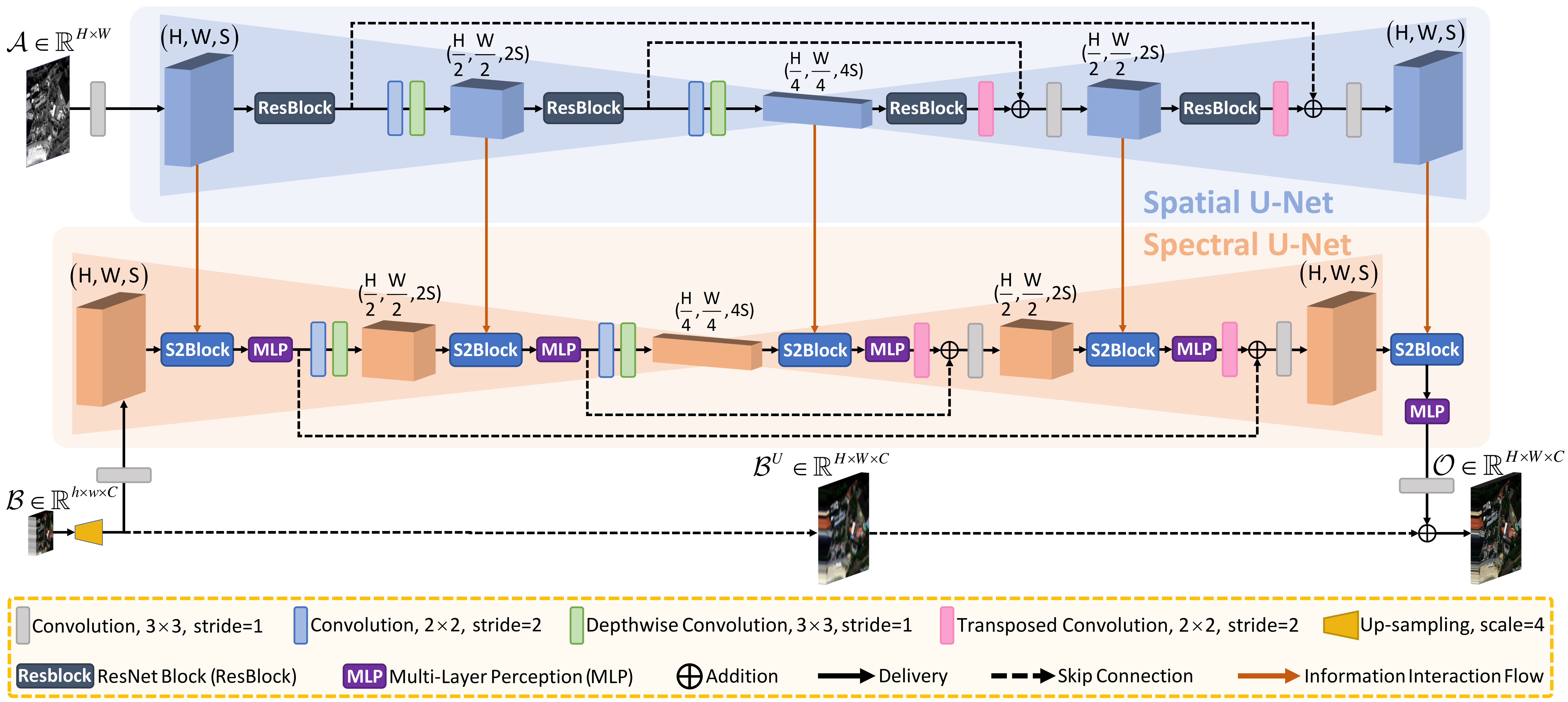}}
			\centering
		\end{minipage}
	\end{center}
	\caption{The overall structure of the proposed method. The U2Net employs a spatial U-Net and a spectral U-Net to extract spatial details and spectral characteristics, respectively. Besides, feature maps from different sources are integrated through the well-designed S2Blocks. The notations in this figure are explained in Section~\ref{s31}. \label{u2net}}
\end{figure*}

\section{Related Work}
\label{s2}
\noindent\textbf{DL-based Methods.} In recent years, a number of DL-based image fusion methods have been proposed. These methods outperform traditional works due to the superior capabilities of DL in feature extraction and nonlinear fitting.
For pansharpening, the pioneering work is the PNN \cite{2016Pansharpening} which utilizes three convolutional layers to achieve the best performance at that time. Since then, impressive methods such as PanNet \cite{8237455}, DiCNN \cite{2018Pansharpening}, and FusionNet \cite{2020Detail} have successively emerged, further validating the potential of DL in the field of pansharpening. There are also many exceptional DL-based works in the field of HISR, including ResTFNet \cite{2018Remote}, SSRNet \cite{9186332}, and Fusformer \cite{9841513}. However, due to unreasonable structural design, most DL-based image fusion approaches suffer from drawbacks such as spectral distortion and poor generalization ability.

\noindent\textbf{U-Net.} The U-shape network is put forward by \cite{ronneberger2015u} for the pixel-wise segmentation problem. It utilizes a series of symmetric down-sampling and up-sampling layers to capture information hierarchically. The U-shape network boasts a strong feature extraction capability, as evidenced by its extensive application in many pixel-level computer vision tasks. 
Recently, U-Net has also been introduced into image fusion tasks, and the representatives include DCFNet \cite{Wu_2021_ICCV} and MUCNN \cite{10.1145/3474085.3475600}. It is worth noting that these methods only employ concatenation to merge images from different sources. Then, the cascaded output is fed into a single-branch U-shape network for feature extraction. This structural design can lead to insufficient information fusion and inefficient feature learning, thus requiring numerous parameters to attain satisfactory outcomes.

\noindent\textbf{Motivation.}
Images obtained from different sensors possess unique properties, \emph{e.g.}, PAN/RGB images exhibit rich spatial details, whereas LRMS/LRHS images contain a wealth of spectral information. Therefore, it is imperative to consider their differences when performing image fusion, which aims to produce a fused outcome from the two types of inputs. However, the majority of previous studies employ a single-branch network to uniformly extract spatial and spectral characteristics, as illustrated in Fig.~\ref{cmprr}. Furthermore, these approaches merely apply concatenation to combine the two kinds of images. Consequently, they suffer from significant issues, such as inefficient feature learning, poor generalization ability, and insufficient information integration. The above situation motivates us to propose U2Net, which captures spatial and spectral features discriminately and hierarchically using two U-shape networks. Besides, the well-designed S2Block enables the effective integration of feature maps from different sources.

\section{The Proposed Method}
\label{s3}
\subsection{Notations}
\label{s31}
The PAN/RGB image is represented as $\mathcal{A} \in \mathbb{R}^{H \times W \times c}$, where $H$, $W$, and $c$ denote height, width, and input channel, respectively. 
$\mathcal{B} \in \mathbb{R}^{h \times w \times C}$ represents the LRMS/LRHS image, in which $h=\frac{H}{4}$ and $w=\frac{W}{4}$. Besides, $C$ denotes the spectral band.
The up-sampled LRMS/LRHS image, desired HRMS/HRHS image, and ground-truth (GT) image are represented as ${\mathcal{B}}^U\in \mathbb{R}^{H \times W \times C}$, $\mathcal{O}\in \mathbb{R}^{H \times W \times C}$, and $\mathcal{X} \in \mathbb{R}^{H \times W \times C}$, respectively.

\begin{figure*}[t]
	\begin{center}
		\begin{minipage}{1\linewidth}
			{\includegraphics[width=0.88\linewidth]{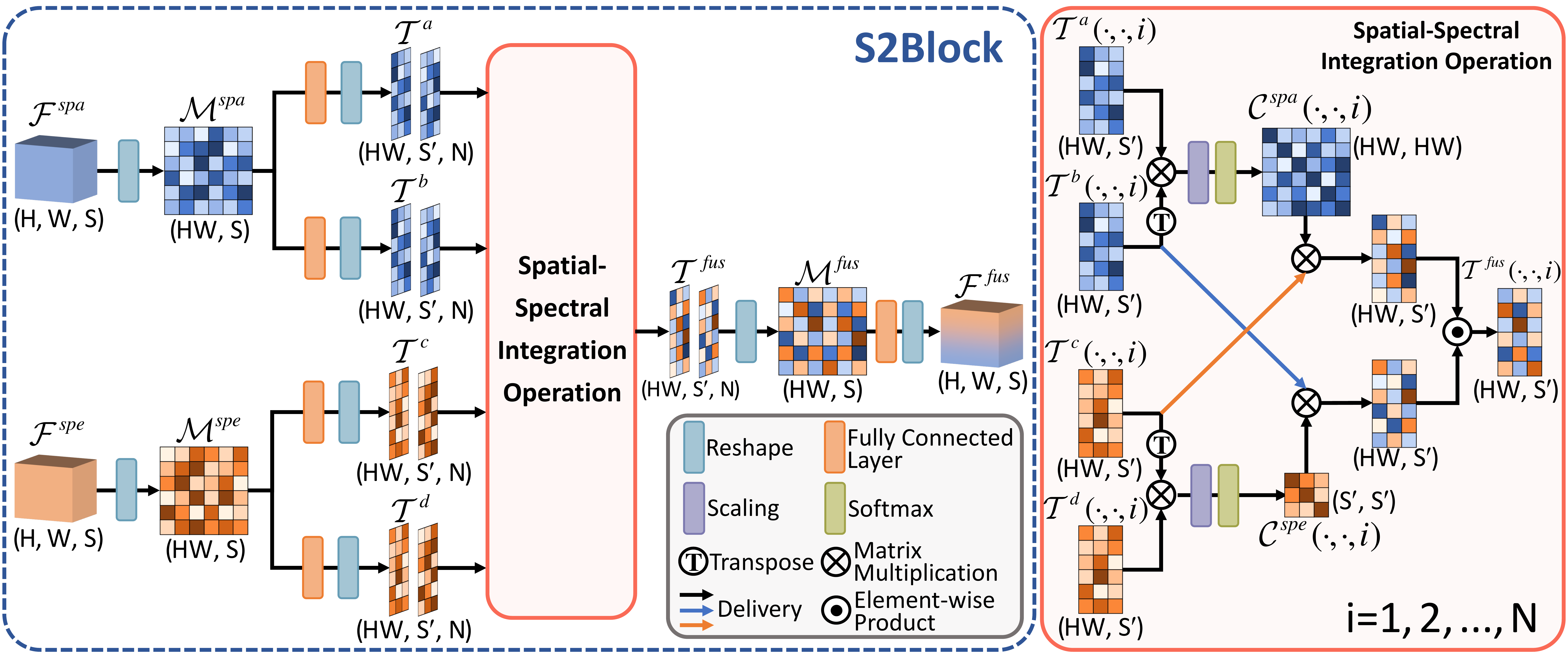}}
			\centering
		\end{minipage}
	\end{center}
	\caption{The structure of S2Block. To enhance the description of spatial details and spectral characteristics, we begin by producing the spatial self-correlation matrices $\mathcal{C}^{spa}$ and spectral self-correlation matrices $\mathcal{C}^{spe}$. After that, a series of operators are employed to integrate spatial and spectral information. The notations in this figure are explained in Section \ref{s33}. \label{s2block}}
\end{figure*}

\subsection{U2Net}
To learn features from diverse images in a discriminative and hierarchical manner, we develop a double U-shape network architecture consisting of a spatial U-Net and a spectral U-Net, as shown in Fig.~\ref{u2net}. The spatial U-Net focuses on extracting spatial details from $\mathcal{A}$, while the spectral U-Net is designed to collect the spectral data in $\mathcal{B}$. In order to capture sufficient deep-level information under limited network parameters, we process features maps at three distinct scales, \emph{i.e.}, $H\times W\times S$ ($S$ denotes the channel number of input feature maps), $\frac{H}{2}\times \frac{W}{2}\times 2S$, and $\frac{H}{4}\times \frac{W}{4}\times 4S$. Thus, the learning process of our U-Net consists of five stages. Each stage employs a neural network to extract information from the feature map of a particular size. According to the structural symmetry of the U-shape network, feature maps of sizes $H\times W\times S$, $\frac{H}{2}\times \frac{W}{2}\times 2S$, and $\frac{H}{4}\times \frac{W}{4}\times 4S$ are processed in stages one and five, stages two and four, and stage three, respectively. Between each pair of adjacent stages, there exists a step that involves operations for down-sampling/up-sampling and dimension transformation. In the initial two steps, we utilize $2\times2$ convolution kernels with a stride of $2$ for down-sampling and apply depth-wise convolutional layers to augment the channel numbers of feature maps. For the final two steps, we use transposed convolutional layers with $2\times2$ kernels and a stride of $2$ to achieve both up-sampling and channel reduction. In addition, the outputs of stages one and two are summed with the inputs of stages five and four, respectively.


This paragraph describes how the spatial U-Net captures spatial details. Firstly, a convolutional layer with $3\times3$ kernels is employed to increase the dimension of $\mathcal{A}$, generating the input for the spatial U-Net. Then, in the initial four stages of the spatial U-shape network, we utilize ResNet blocks (ResBlock) to extract spatial information. Each ResBlock comprises two convolutional layers with $3\times3$ kernels and a layer of leaky rectified linear units (LReLU). To avoid gradient disappearance, a skip connection is established between the input and output. 


This paragraph explains how the spectral U-Net is utilized to construct $\mathcal{O}$. Firstly, we up-sample $\mathcal{B}$ to obtain ${\mathcal{B}}^U$ with high spatial resolution. Next, we employ $3\times3$ convolution kernels to augment the channel number of ${\mathcal{B}}^U$, producing the input for the spectral U-Net. At each stage of the spectral U-shape network, the spectral feature map is combined with the spatial one using S2Block, resulting in a fused outcome. Then, we apply the multi-layer perception (MLP) to acquire spectral characteristics from the fused outcome while preserving spatial information. The MLP primarily consists of two fully connected layers and an LReLU layer. Similar to ResBlock, the input is linked to the output. 
Upon obtaining the output of the final stage, we utilize a convolutional layer with $3\times3$ kernels to reconstruct it into a feature map of the size $H\times W\times C$. The feature map is then added to ${\mathcal{B}}^U$, generating the desired $\mathcal{O}$.

\begin{table*}[t]	
	\centering\renewcommand\arraystretch{1.4}\setlength{\tabcolsep}{10.5pt}
	\footnotesize
	\caption{Quantitative results on 20 reduced-resolution and 20 full-resolution samples of WV3. (\textbf{\textcolor{red}{Red}}: best; \textcolor{blue}{Blue}: second best).}
	\begin{tabular}{ccccc|ccc}
		\toprule
		
		\multirow{2}*{\textbf{Method}}
		& \multicolumn{4}{c}{\textbf{Reduced-Resolution}} &\multicolumn{3}{c}{\textbf{Full-Resolution}}\\
		
		\Xcline{2-8}{0.4pt}
		
		&\multicolumn{1}{c}{PSNR($\pm$std)} & \multicolumn{1}{c}{Q8($\pm$std)} &\multicolumn{1}{c}{SAM($\pm$std)} &\multicolumn{1}{c} {ERGAS($\pm$std)} &\multicolumn{1}{c}{$\rm{D_{\lambda}}$($\pm$std)} &\multicolumn{1}{c}{$\rm{D_{s}}$($\pm$std)} &\multicolumn{1}{c}{QNR($\pm$std)} \\
		
		\midrule
		
		\textbf{BT-H} \cite{aiazzi2006mtf} & 33.080$\pm$2.880 & 0.832$\pm$0.094 & 4.920$\pm$1.425 & 4.580$\pm$1.496          
		& 0.0574$\pm$0.0232  & 0.0810$\pm$0.0374 & 0.8670$\pm$0.0540\\ 
		\textbf{TV} \cite{palsson2013new} & 32.381$\pm$2.328 & 0.795$\pm$0.120 & 5.692$\pm$1.808 & 4.855$\pm$1.434
		& 0.0234$\pm$0.0061  & 0.0393$\pm$0.0227 & 0.9383$\pm$0.0269\\ 
		\textbf{MTF-GLP-HPM} \cite{6616569} & 33.095$\pm$2.800 & 0.835$\pm$0.092 & 5.333$\pm$1.761 & 4.616$\pm$1.503
		& 0.0206$\pm$0.0082  & 0.0630$\pm$0.0284 & 0.9180$\pm$0.0346\\ 
		\textbf{MTF-GLP-FS} \cite{vivone2018full} & 32.963$\pm$2.753 & 0.833$\pm$0.092 & 5.315$\pm$1.765 & 4.700$\pm$1.597
		& 0.0197$\pm$0.0078  & 0.0630$\pm$0.0289 & 0.9187$\pm$0.0347\\ 
		\textbf{BDSD-PC} \cite{2019Robust} & 32.970$\pm$2.784 & 0.829$\pm$0.097 & 5.428$\pm$1.822 & 4.697$\pm$1.617
		& 0.0625$\pm$0.0235  & 0.0730$\pm$0.0356 & 0.8698$\pm$0.0531\\ 
		\textbf{PNN} \cite{2016Pansharpening} & 37.313$\pm$2.646 & 0.893$\pm$0.092 & 3.677$\pm$0.762 & 2.681$\pm$0.647
		& 0.0213$\pm$0.0080  & 0.0428$\pm$0.0147 & 0.9369$\pm$0.0212\\ 
		\textbf{PanNet} \cite{8237455} & 37.346$\pm$2.688 & 0.891$\pm$0.093 & 3.613$\pm$0.766 & 2.664$\pm$0.688
		& \textbf{\textcolor{red}{0.0165}}$\pm$0.0074  & 0.0470$\pm$0.0210 & 0.9374$\pm$0.0271\\ 
		\textbf{MSDCNN} \cite{8127731} & 37.068$\pm$2.686 & 0.890$\pm$0.090 & 3.777$\pm$0.803 & 2.760$\pm$0.689
		& 0.0230$\pm$0.0091 & 0.0467$\pm$0.0199 & 0.9316$\pm$0.0271\\
		\textbf{DiCNN} \cite{2018Pansharpening} & 37.390$\pm$2.761 & 0.900$\pm$0.087 & 3.592$\pm$0.762 & 2.672$\pm$0.662
		& 0.0362$\pm$0.0111  & 0.0462$\pm$0.0175 & 0.9195$\pm$0.0258  \\
		
		\textbf{BDPN} \cite{zhang2019pan} & 36.191$\pm$2.702 & 0.871$\pm$0.100 & 4.201$\pm$0.857 & 3.046$\pm$0.732
		& 0.0364$\pm$0.0142  & 0.0459$\pm$0.0192 & 0.9196$\pm$0.0308\\ 
		\textbf{FusionNet} \cite{2020Detail} & 38.047$\pm$2.589 & 0.904$\pm$0.090 & 3.324$\pm$0.698 & 2.465$\pm$0.644
		& 0.0239$\pm$0.0090  & 0.0364$\pm$0.0137 & 0.9406$\pm$0.0197\\   
		\textbf{MUCNN} \cite{10.1145/3474085.3475600} & 38.262$\pm$2.703 & 0.911$\pm$0.089 & 3.206$\pm$0.681 & 2.400$\pm$0.617
		& 0.0258$\pm$0.0111  & \textcolor{blue}{0.0327}$\pm$0.0140 & \textcolor{blue}{0.9424}$\pm$0.0205\\   
		\textbf{LAGNet} \cite{jin2022aaai} & 38.592$\pm$2.778 & 0.910$\pm$0.091 & 3.103$\pm$0.558 & \textcolor{blue}{2.292}$\pm$0.607
		& 0.0368$\pm$0.0148  & 0.0418$\pm$0.0152& 0.9230$\pm$0.0247\\  
		\textbf{PMACNet} \cite{9764690} & \textcolor{blue}{38.595}$\pm$2.882 & \textcolor{blue}{0.912}$\pm$0.092 & \textcolor{blue}{3.073}$\pm$0.623 & 2.293$\pm$0.532
		& 0.0540$\pm$0.0232  & 0.0336$\pm$0.0115 & 0.9143$\pm$0.0281\\  	\textbf{U2Net} & \textbf{\textcolor{red}{39.117}}$\pm$3.009 & \textbf{\textcolor{red}{0.920}}$\pm$0.085 & \textbf{\textcolor{red}{2.888}}$\pm$0.581 & \textbf{\textcolor{red}{2.149}}$\pm$0.525
		& \textcolor{blue}{0.0178}$\pm$0.0072  & \textbf{\textcolor{red}{0.0313}}$\pm$0.0075 & \textbf{\textcolor{red}{0.9514}}$\pm$0.0115\\       \midrule
		
		\textbf{Ideal value} 
		&\multicolumn{1}{c}{\textbf{+$\infty$}}
		&\multicolumn{1}{c}{\textbf{\textbf{1}}}
		&\multicolumn{1}{c}{\textbf{\textbf{0}}}
		&\multicolumn{1}{c}{\textbf{\textbf{0}}}
		&\multicolumn{1}{c}{\textbf{\textbf{0}}}
		&\multicolumn{1}{c}{\textbf{\textbf{0}}}
		&\multicolumn{1}{c}{\textbf{\textbf{1}}}
		\\ 
		\bottomrule
	\end{tabular}
	\label{rr}	
\end{table*}

\subsection{S2Block}
\label{s33}
Most existing image fusion approaches apply concatenation to integrate spatial and spectral information. However, this operation disregards the distinctions between the two types of images, resulting in unsatisfactory fusion outcomes. To overcome this limitation, we develop a novel structure called S2Block, as illustrated in Fig.~\ref{s2block}, for effective spatial-spectral integration. Next, we will take the S2Block in the first stage of the spectral U-Net as an example to explain this structure.


The spatial and spectral feature maps input to the S2Block are denoted as $\mathcal{F}^{spa}\in \mathbb{R}^{{H} \times{W}\times {S}}$ and $\mathcal{F}^{spe} \in \mathbb{R}^{{H} \times{W}\times {S}}$. For $\mathcal{F}^{spa}$, we first reshape it into a matrix denoted $\mathcal{M}^{spa}\in \mathbb{R}^{HW \times S}$, with each row representing the feature vector of a particular spatial location. Then, the $\mathcal{M}^{spa}$ is simultaneously processed by two parallel fully connected layers, producing two matrices of the same size. To better utilize the information on feature vectors, we divide each matrix evenly into several smaller parts by column. Technically, the matrices are reshaped into two tensors, denoted as $\mathcal{T}^{a}\in \mathbb{R}^{{HW}\times{S'}\times{N}}$ and $\mathcal{T}^{b}\in \mathbb{R}^{{HW}\times{S'}\times{N}}$, where $N$ is the number of small parts and $S'=\frac{S}{N}$. For $\mathcal{F}^{spe}$, we first reshape it into a matrix denoted $\mathcal{M}^{spe}\in \mathbb{R}^{HW \times S}$. Similarly, the $\mathcal{M}^{spe}$ is transformed into two tensors, denoted as $\mathcal{T}^{c}\in \mathbb{R}^{{HW}\times{S'}\times{N}}$ and $\mathcal{T}^{d}\in \mathbb{R}^{{HW}\times{S'}\times{N}}$.

This paragraph explains the spatial-spectral integration operation (SSIO), which is the core of S2Block. In SSIO, we first produce two sets of square matrices, namely \emph{spatial self-correlation matrices} and \emph{spectral self-correlation matrices}, to accurately depict spatial details and spectral characteristics. Next, a series of operators are utilized to merge the square matrices with the spatial and spectral data, resulting in a fusion outcome. Specifically, we represent the set of spatial self-correlation matrices as $\mathcal{C}^{spa} \in \mathbb{R}^{HW\times HW\times N}$, and the production of its $i^{th}$ square matrix is expressed as:
\begin{equation}
	{\mathcal{C}^{spa}}{(\cdot, \cdot, i)} = {\rm{Softmax}}\Big(\frac{{\mathcal{T}^{a}}{(\cdot, \cdot, i)}\{{\mathcal{T}^{b}}{(\cdot, \cdot, i)}\}^T}{\sqrt{S'}}\Big),
\end{equation} 
where ${\mathcal{C}^{spa}}{(\cdot, \cdot, i)}\in \mathbb{R}^{HW\times HW}$ denotes the $i^{th}$ matrix of $\mathcal{C}^{spa}$. ${\mathcal{T}^{a}}{(\cdot, \cdot, i)} \in \mathbb{R}^{HW\times S'}$ and ${\mathcal{T}^{b}}{(\cdot, \cdot, i)} \in \mathbb{R}^{HW\times S'}$ represent the $i^{th}$ matrices in ${\mathcal{T}^{a}}$ and ${\mathcal{T}^{b}}$, respectively. Besides, $T$ defines the transpose operation and $\rm{Softmax(\cdot)}$ stands for the Softmax function. The spatial self-correlation matrices offer a concrete and intuitive representation of spatial information, as each value of ${\mathcal{C}^{spa}}{(\cdot, \cdot, i)}$ signifies the similarity between two spatial locations in $\mathcal{A}$. The set of spectral self-correlation matrices is represented as $\mathcal{C}^{spe} \in \mathbb{R}^{S'\times S'\times N}$, and we express its $i^{th}$ matrix as:
\begin{equation}
	{\mathcal{C}^{spe}}{(\cdot, \cdot, i)} = {\rm{Softmax}}\Big(\frac{\{{\mathcal{T}^{c}}{(\cdot, \cdot, i)}\}^T{\mathcal{T}^{d}}{(\cdot, \cdot, i)}}{\frac{\sqrt{(S')^{3}}}{HW}}\Big),
\end{equation} 
where ${\mathcal{C}^{spe}}{(\cdot, \cdot, i)}\in \mathbb{R}^{S'\times S'}$ denotes the $i^{th}$ square matrix of $\mathcal{C}^{spe}$. ${\mathcal{T}^{c}}{(\cdot, \cdot, i)} \in \mathbb{R}^{HW\times S'}$ and ${\mathcal{T}^{d}}{(\cdot, \cdot, i)} \in \mathbb{R}^{HW\times S'}$ stand for the $i^{th}$ matrices in ${\mathcal{T}^{c}}$ and ${\mathcal{T}^{d}}$. Since each value of ${\mathcal{C}^{spe}}{(\cdot, \cdot, i)}$ represents the similarity between two channels of $\mathcal{B}$, the spectral self-correlation matrices provide a tangible and intuitive description of spectral characteristics. Upon obtaining the ${\mathcal{C}^{spa}}{(\cdot, \cdot, i)}$ and ${\mathcal{C}^{spe}}{(\cdot, \cdot, i)}$, we combine them with the spatial and spectral data, expressed as:
\begin{equation}\label{mmmm}
	{\mathcal{T}^{fus}}(\cdot,\cdot,i) = \{\mathcal{C}^{spa}(\cdot,\cdot,i)\mathcal{T}^{c}(\cdot, \cdot, i)\}\odot\{(\mathcal{T}^{b}(\cdot, \cdot, i)\mathcal{C}^{spe}(\cdot,\cdot,i)\},
\end{equation} 
where $\mathcal{T}^{fus} \in \mathbb{R}^{HW\times S'\times N}$ denotes the fused output that contains both spatial and spectral information. $\mathcal{T}^{fus}(\cdot, \cdot, i) \in \mathbb{R}^{HW\times S'}$ represents the $i^{th}$ matrix in $\mathcal{T}^{fus}$. Additionally, $\odot$ defines the element-wise multiplication. Compared with other fusion techniques like concatenation, the SSIO enables effective and comprehensive integration of spatial details and spectral characteristics.

After acquiring $\mathcal{T}^{fus}$, we reshape it into a fusion matrix, denoted as $\mathcal{M}^{{fus}} \in \mathbb{R}^{{HW}\times{S}}$. Subsequently, we employ a fully connected layer to process the $\mathcal{M}^{fus}$ and convert it into a spatial-spectral-integrated feature map, represented as $\mathcal{F}^{fus} \in \mathbb{R}^{{H} \times{W}\times {S}}$.

Besides, please refer to the \emph{Sup. Mat.} for a comprehensive explanation regarding the relationship between our S2Block and multi-head attention in Transformer \cite{vaswani2017attention}.


\subsection{Loss Function}
The main contributions of this work focus on the network architecture, thus we only employ the commonly used $\ell_{1}$ loss function for network training, shown as follows:
\begin{equation}\label{loss}
	{\mathcal{L}oss} =\frac{1}{M}\sum_{m=1}^{M}{\| f_\Theta({\mathcal{A}^{\{m\}},\mathcal{B}^{\{m\}}}) - \mathcal{X}^{\{m\}}\|}_{1},
\end{equation} 
where $\mathcal{A}^{\{m\}}$, $\mathcal{B}^{\{m\}}$, and $\mathcal{X}^{\{m\}}$ represent the $m^{th}$ PAN/RGB image, LRMS/LRHS image, and GT image in the training dataset. $f_\Theta(\cdot)$ denotes the U2Net with learnable parameters $\Theta$, and $M$ is the total number of training examples. Besides, $\|\cdot\|_1$ defines the $\ell_{1}$ norm.

\begin{figure*}[t]
	\begin{center}
		\begin{minipage}[t]{0.95\linewidth}
			\begin{minipage}[t]{0.121\linewidth}
				{\includegraphics[width=1\linewidth]{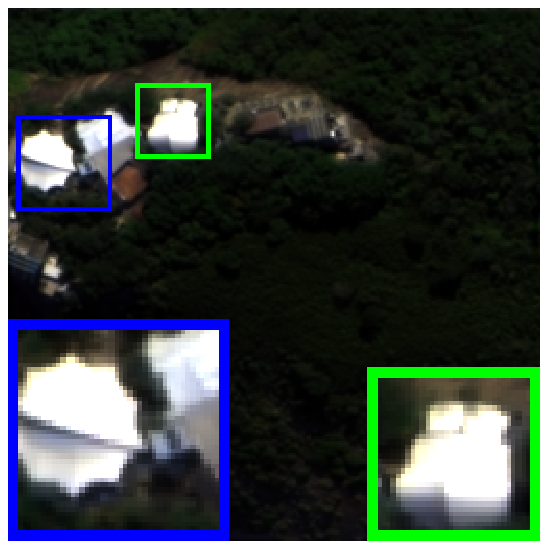}}
				{\includegraphics[width=1\linewidth]{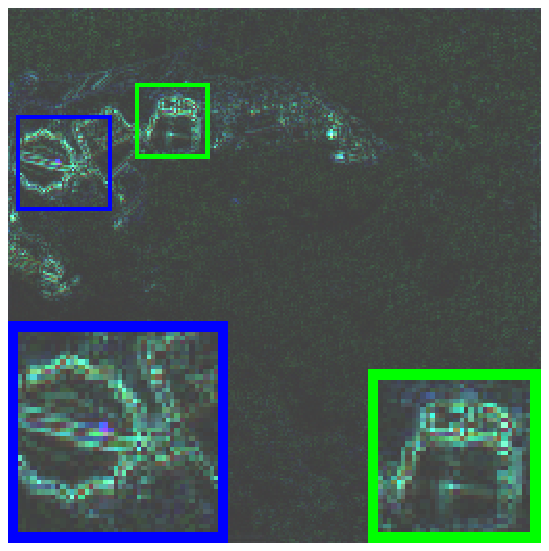}}
				\vspace{2pt}
				{BT-H}
				{\includegraphics[width=1\linewidth]{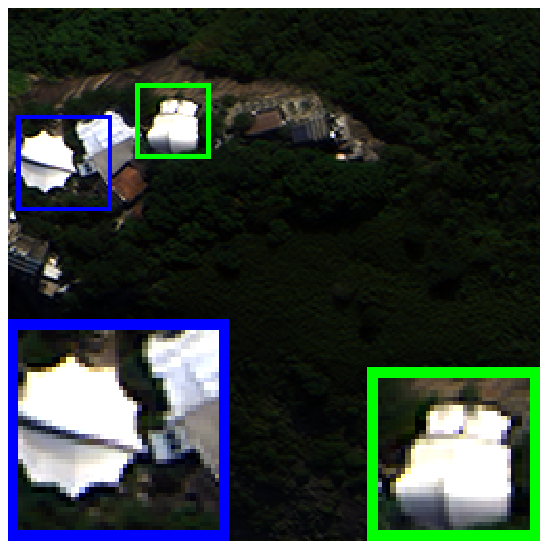}}
				{\includegraphics[width=1\linewidth]{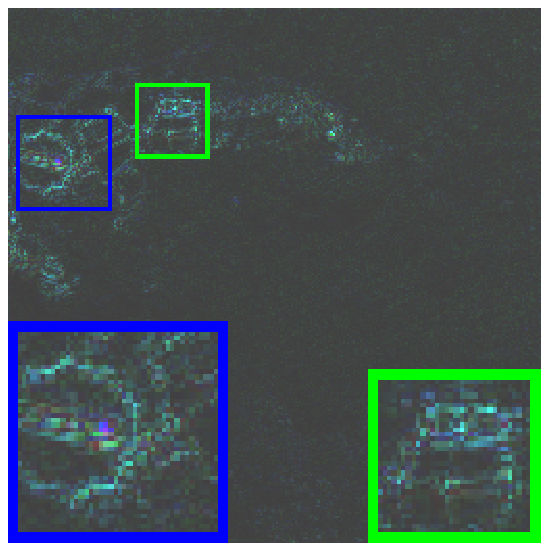}}
				{DiCNN}
				\centering
				
			\end{minipage}
			\begin{minipage}[t]{0.121\linewidth}
				{\includegraphics[width=1\linewidth]
					{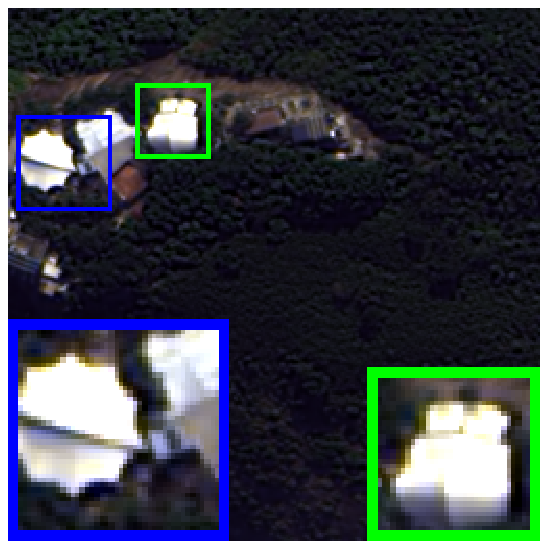}}
				{\includegraphics[width=1\linewidth]{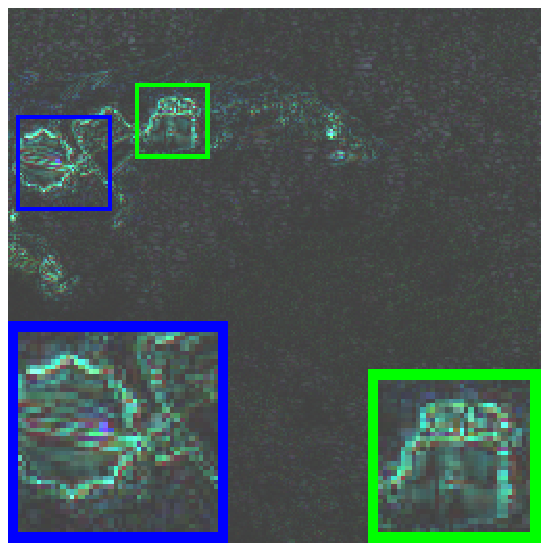}}
				\vspace{2pt}
				{TV}
				{\includegraphics[width=1\linewidth]{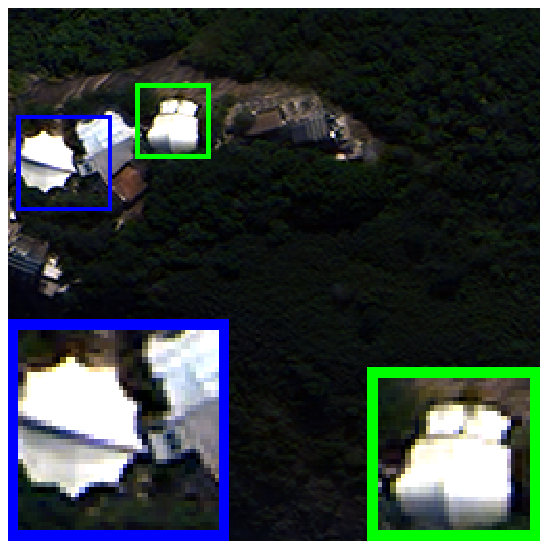}}
				{\includegraphics[width=1\linewidth]{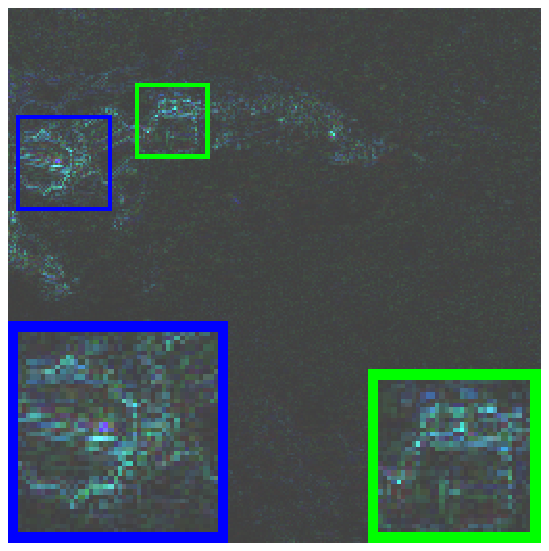}}
				{BDPN}
				\centering
				
			\end{minipage}
			\begin{minipage}[t]{0.121\linewidth}
				{\includegraphics[width=1\linewidth]{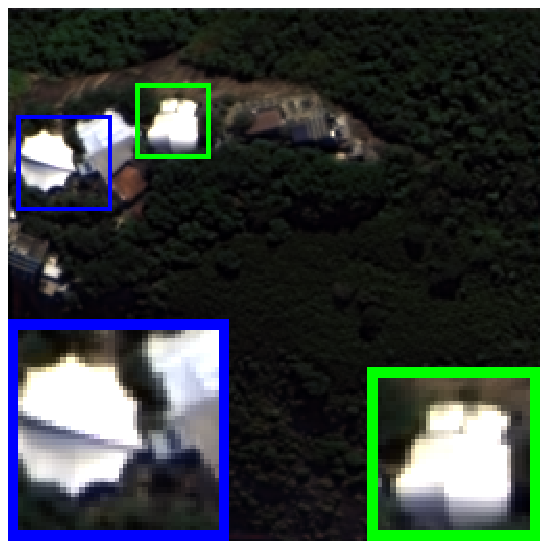}}
				{\includegraphics[width=1\linewidth]{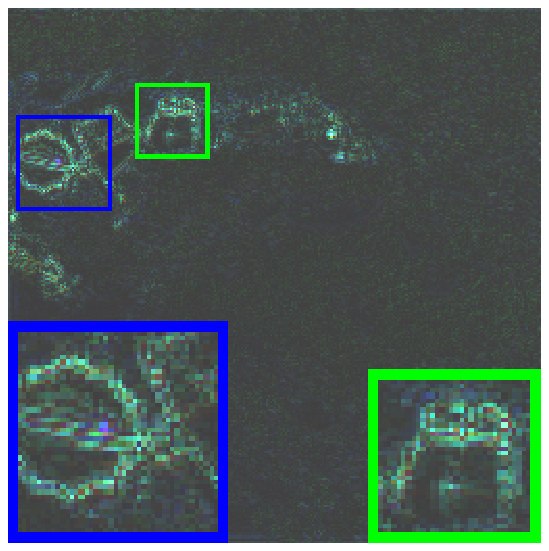}}
				\vspace{2pt}
				{MTF-GLP-HPM}
				{\includegraphics[width=1\linewidth]{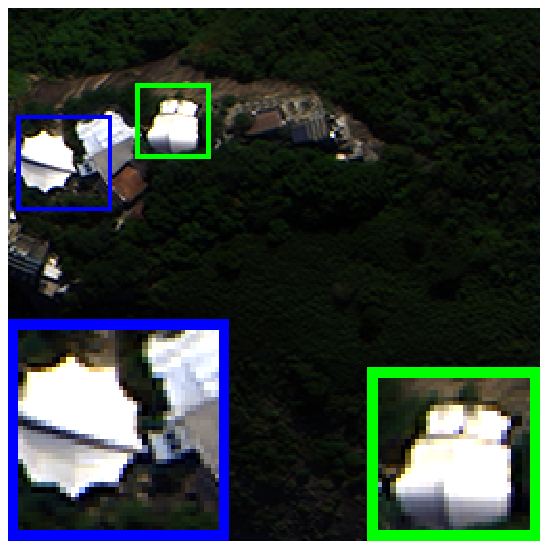}}
				{\includegraphics[width=1\linewidth]{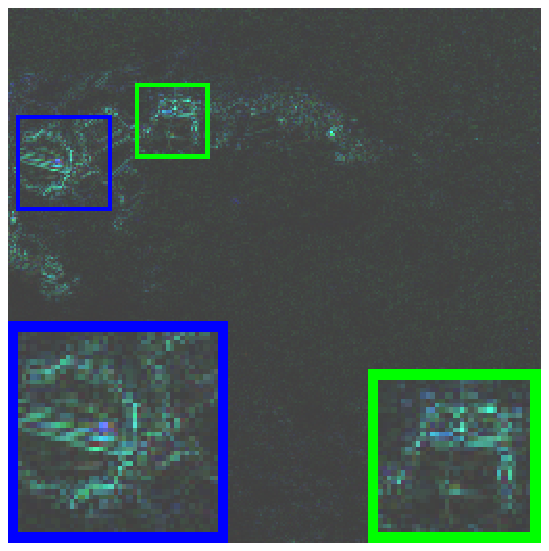}}
				{FusionNet}
				\centering
				
			\end{minipage}
			\begin{minipage}[t]{0.121\linewidth}
				{\includegraphics[width=1\linewidth]{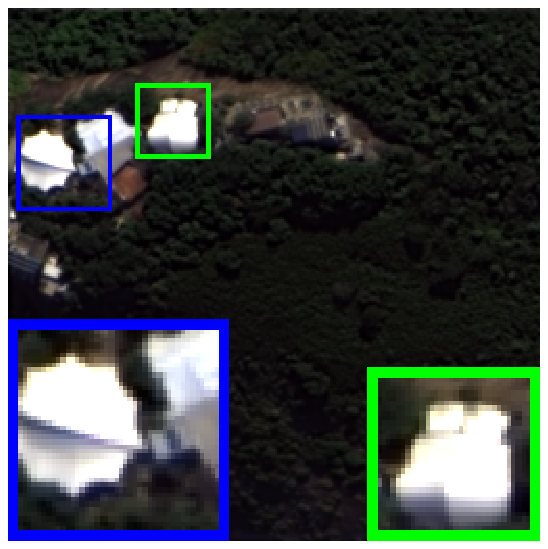}}
				{\includegraphics[width=1\linewidth]{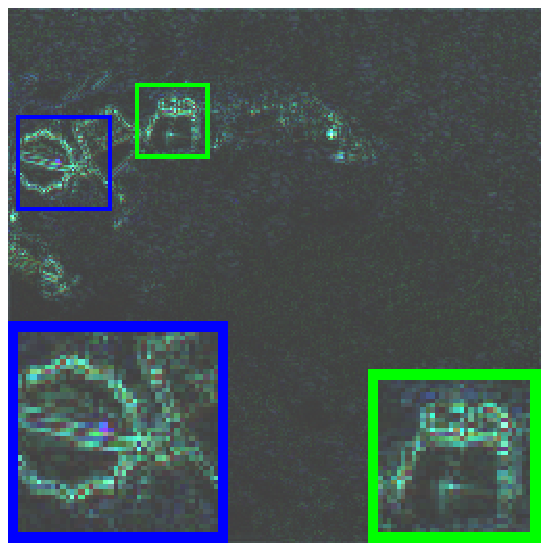}}
				\vspace{2pt}
				{MTF-GLP-FS}
				{\includegraphics[width=1\linewidth]{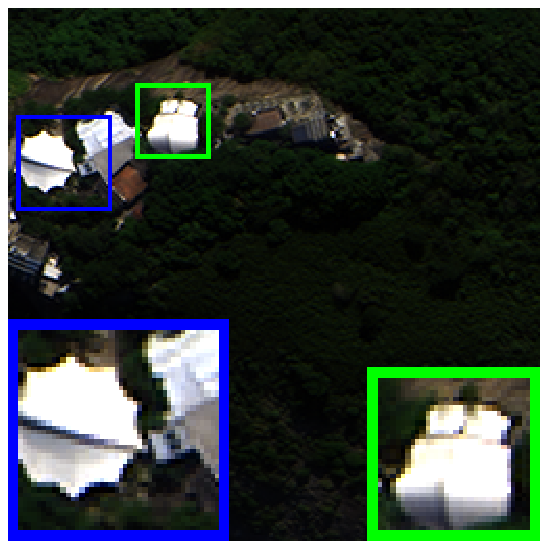}}
				{\includegraphics[width=1\linewidth]{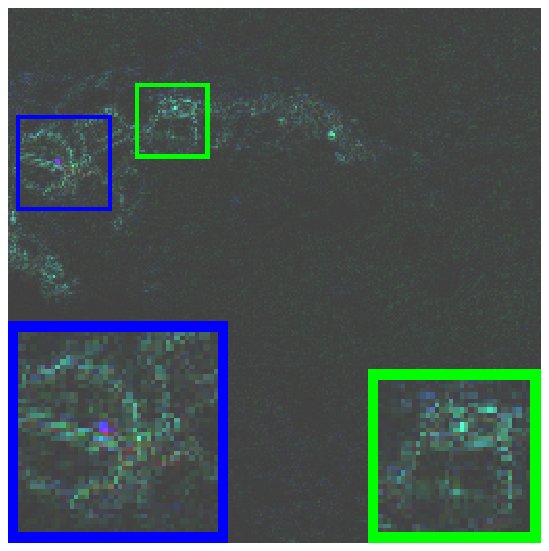}}
				{MUCNN}
				\centering
				
			\end{minipage}
			\begin{minipage}[t]{0.121\linewidth}
				{\includegraphics[width=1\linewidth]{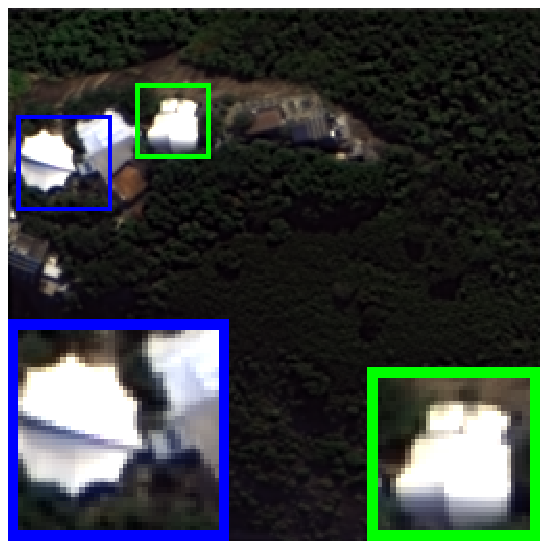}}
				{\includegraphics[width=1\linewidth]{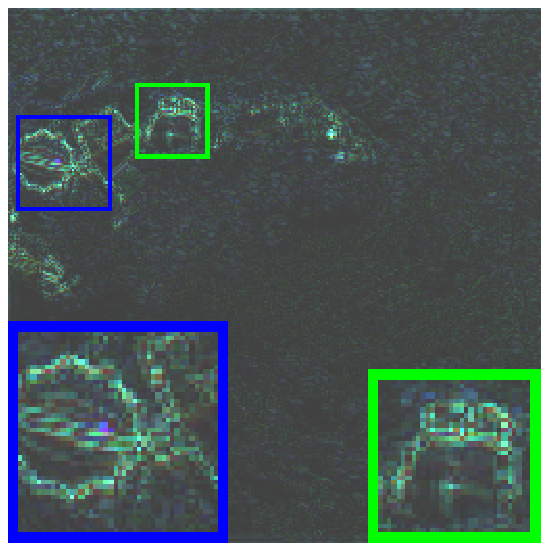}}
				\vspace{2pt}
				{BDSD-PC}
				{\includegraphics[width=1\linewidth]{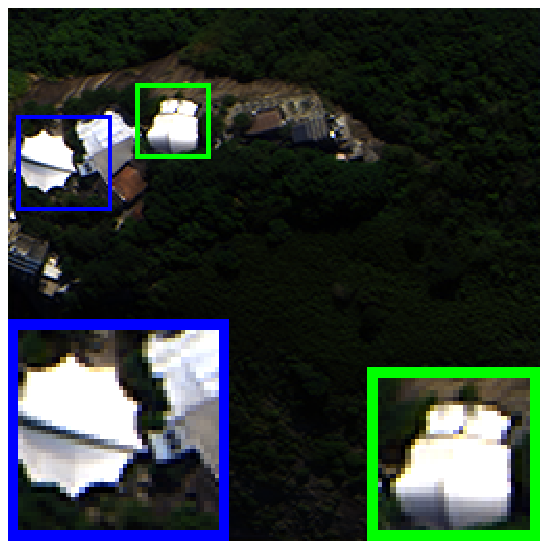}}
				{\includegraphics[width=1\linewidth]{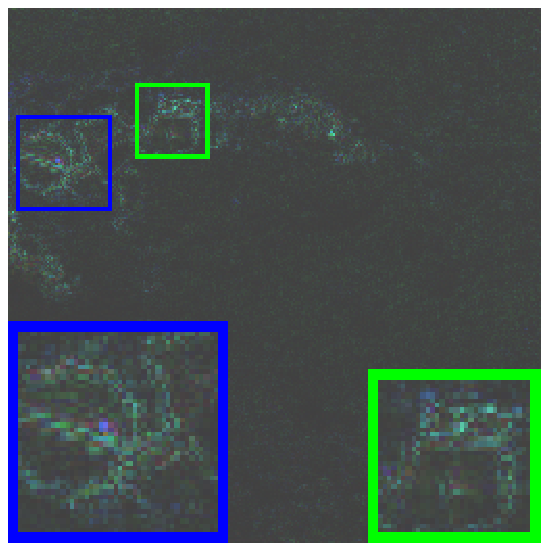}}
				{LAGNet}
				\centering
				
			\end{minipage}
			\begin{minipage}[t]{0.121\linewidth}
				{\includegraphics[width=1\linewidth]{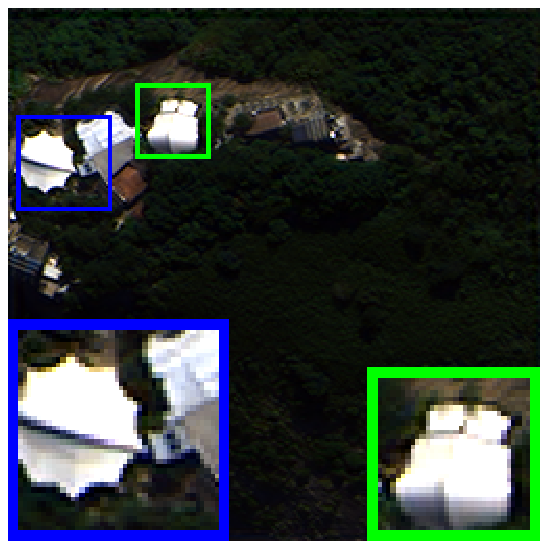}}
				{\includegraphics[width=1\linewidth]{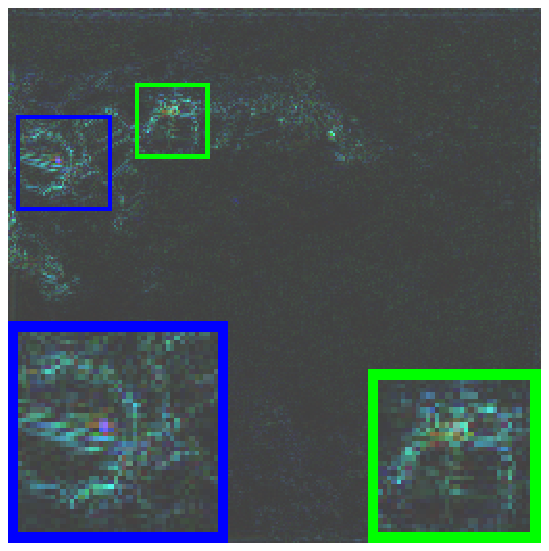}}
				\vspace{2pt}
				{PNN}
				{\includegraphics[width=1\linewidth]{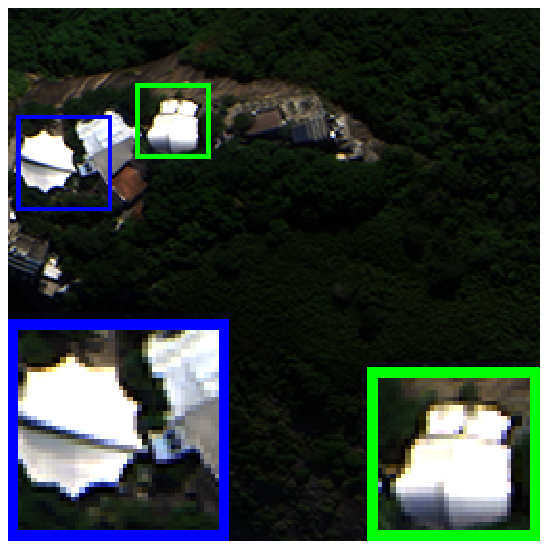}}
				{\includegraphics[width=1\linewidth]{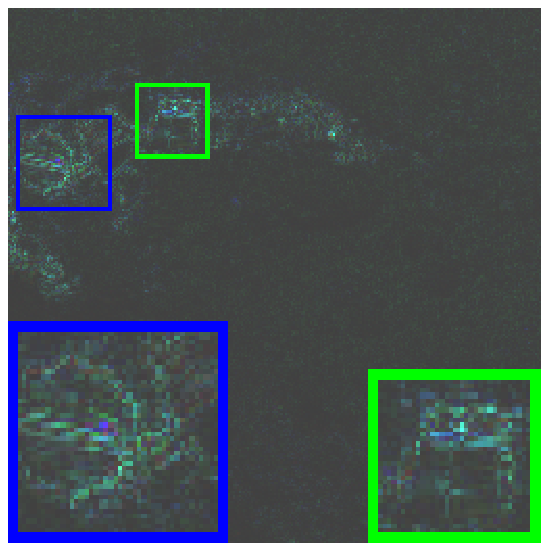}}
				{PMACNet}
				\centering
				
			\end{minipage}
			\begin{minipage}[t]{0.121\linewidth}
				{\includegraphics[width=1\linewidth]{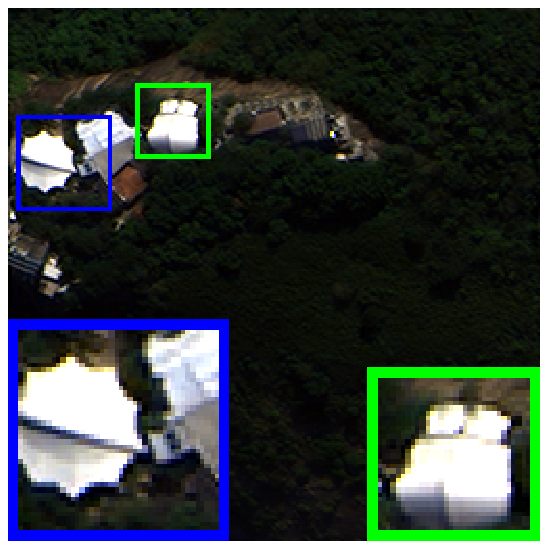}}
				{\includegraphics[width=1\linewidth]{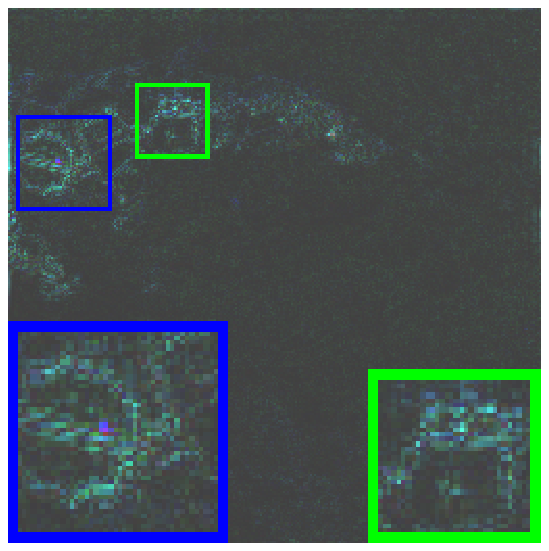}}
				\vspace{2pt}
				{PanNet}
				{\includegraphics[width=1\linewidth]{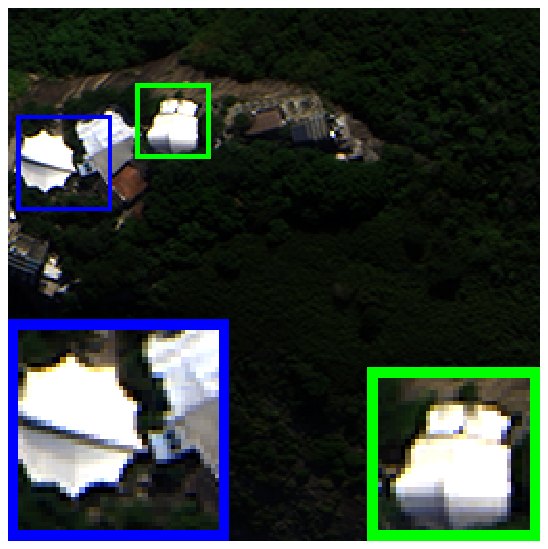}}
				{\includegraphics[width=1\linewidth]{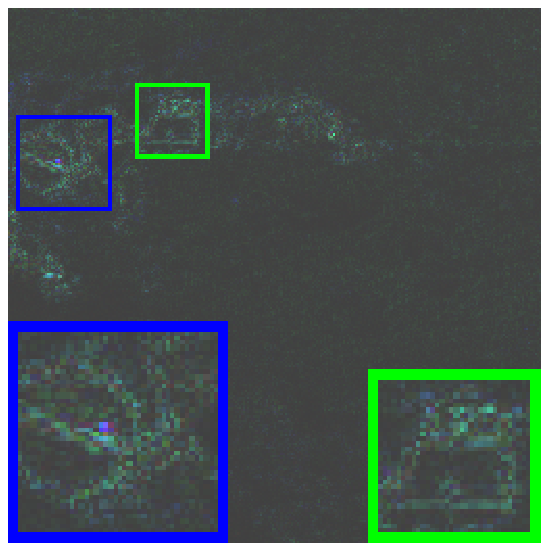}}
				{U2Net}
				\centering
				
			\end{minipage}
			\begin{minipage}[t]{0.121\linewidth}
				{\includegraphics[width=1\linewidth]{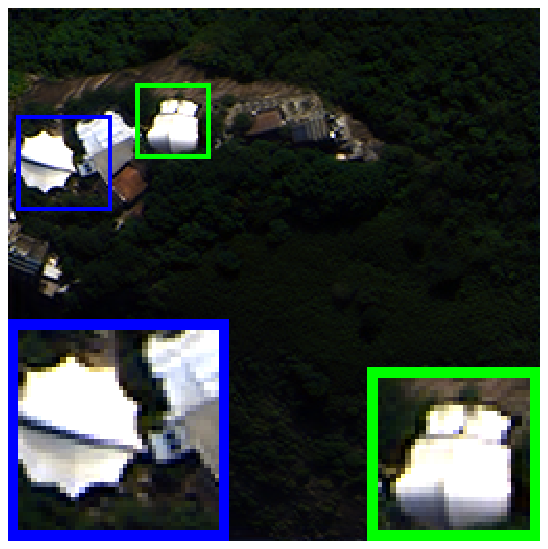}}
				{\includegraphics[width=1\linewidth]{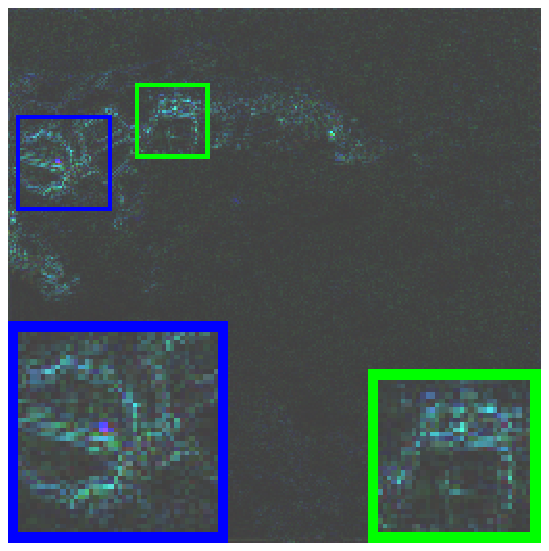}}
				\vspace{2pt}
				{MSDCNN}
				{\includegraphics[width=1\linewidth]{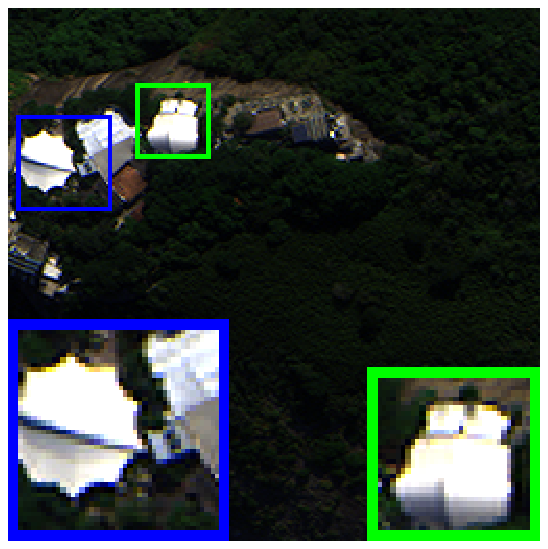}}
				{\includegraphics[width=1\linewidth]{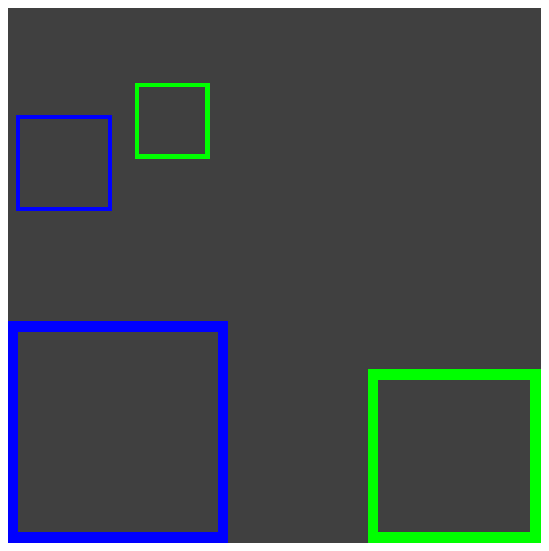}}
				{GT}
				\centering
				
			\end{minipage}
		\end{minipage}
	\end{center}
	\caption{Qualitative evaluation results on a reduced-resolution sample acquired by WV3. The natural color maps are presented in the first and third rows, while the corresponding AEMs are listed in the second and fourth rows.\label{comparison1}}
\end{figure*}

\section{Experiments for Pansharpening}
\label{s4}
To demonstrate the effectiveness of our method, we conduct a series of experiments on datasets acquired by WorldView-3 (WV3) and WorldView-2 (WV2) satellites. The U2Net is compared with several recent SOTA pansharpening approaches.

\subsection{Experiment Settings}
\noindent\textbf{Datasets.} For the pansharpening problem, we train the DL-based methods on a dataset acquired by WV3 which contains 10000 samples (90\% for training and 10\% for validation). Each sample consists of a PAN/LRMS/GT image pair of sizes $64\times64$, $16\times16\times8$, and $64\times64\times8$, respectively. The PAN images have a spatial resolution of 0.3m, while the LRMS images have a spatial resolution of 1.2m. Additionally, the LRMS bands comprise four standard colors (RGB and near-infrared 1) and four new bands (coastal, yellow, red edge, and near-infrared 2). We compare our U2Net with representative pansharpening approaches using various datasets acquired by WV3 and WV2. The testing datasets are categorized into two classes, \emph{i.e.}, the reduced-resolution datasets and the full-resolution datasets. The former includes PAN/LRMS/GT image pairs with dimensions $256\times256$, $64\times64\times8$, and $256\times256\times8$, while the latter consists of PAN/LRMS image pairs of sizes $512\times512$ and $128\times128\times8$. All datasets used in this section are from the PanCollection proposed by \cite{9844267}. The PanCollection offers multiple pansharpening datasets, accompanied by detailed descriptions of data simulation, and can be downloaded from this website\footnote{\url{https://github.com/liangjiandeng/PanCollection}}.

\noindent\textbf{Benchmarks.} 
We compare our method with recent SOTA works consisting of five traditional approaches: BT-H \cite{aiazzi2006mtf}, TV \cite{palsson2013new}, MTF-GLP-HPM \cite{6616569}, MTF-GLP-FS \cite{vivone2018full}, and BDSD-PC \cite{2019Robust}; and nine DL-based methods: PNN \cite{2016Pansharpening}, PanNet \cite{8237455}, MSDCNN \cite{8127731}, DiCNN \cite{2018Pansharpening}, BDPN \cite{zhang2019pan}, FusionNet \cite{2020Detail}, MUCNN \cite{10.1145/3474085.3475600}, LAGNet \cite{jin2022aaai}, and PMACNet \cite{9764690}. To ensure fairness, we train DL-based methods using the same Nvidia GPU-3090 and PyTorch environment.

\noindent\textbf{Evaluation Metrics.} 
Following the research standard of pansharpening, we utilize four metrics to evaluate the results on reduced-resolution datasets, including PSNR, Q8 \cite{2009Hypercomplex}, SAM, and ERGAS \cite{2002Data}.
As for full-resolution datasets, we apply $D_{\lambda}$, $D_{s}$, and QNR indexes \cite{6998089} for evaluation.

\noindent\textbf{Parameters Tuning.} 
For the pansharpening task, we set the values of $S$ and $S'$ in our network to 32 and 16, respectively. Additionally, the value of N depends on the $S$ and $S'$.
On training the U2Net, the initial learning rate, epoch, and batch size are set to 0.001, 360, and 16, respectively. We select Adam as the optimizer, and the learning rate is reduced by half every 100 epochs. As for other DL-based methods, we utilize the default settings in related papers or codes to train the networks.

\subsection{Results on WV3 Datasets}
\noindent\textbf{Reduced-Resolution Assessment.} We assess the performances of representative approaches and our method, using 20 reduced-resolution samples acquired by WV3. The quantitative evaluation outcomes are presented in Tab.~\ref{rr}, and the proposed method obtains the best average results on all quality indexes. Additionally, the qualitative evaluation outcomes on one of the 20 samples are shown in Fig.~\ref{comparison1}, alongside the GT. As the darker absolute error map (AEM) indicates a better result, our U2Net outperforms other approaches. The experimental outcomes above demonstrate that our method is superior to recent SOTA pansharpening works.

\noindent\textbf{Full-Resolution Assessment.} To prove the practical usefulness of our method, we conduct experiments on 20 full-resolution samples acquired by WV3. The quantitative evaluation results are presented in Tab.~\ref{rr}. The U2Net achieves the best overall performance, proving the high application value of our method.

\begin{table}[t]	
	\centering\renewcommand\arraystretch{1.4}\setlength{\tabcolsep}{6pt}
	\footnotesize
	\caption{Quantitative evaluation results of DL-based methods on 20 reduced-resolution samples acquired by WV2. Section \ref{s43} explains the unsatisfactory outcomes of the PMACNet. (\textcolor{red}{Red}: best; \textcolor{blue}{Blue}: second best).}
	\begin{tabular}{ccccccc}
		\toprule
		\textbf{Method} & PSNR($\pm$std)& Q8($\pm$std) & SAM($\pm$std) & ERGAS($\pm$std)\\ 
		\midrule
		\textbf{PNN} & 28.045$\pm$1.865&0.762$\pm$0.093&7.115$\pm$1.682&5.615$\pm$0.943 \\ 
		\textbf{PanNet} & \textcolor{blue}{30.276}$\pm$2.290 & \textcolor{blue}{0.840}$\pm$0.080& \textcolor{blue}{5.495}$\pm$0.713 & \textcolor{blue}{4.337}$\pm$0.520 \\ 
		\textbf{DiCNN} & 27.200$\pm$2.327&0.721$\pm$0.075&6.921$\pm$0.788&6.251$\pm$0.574\\ 
		\textbf{MSDCNN} & 29.441$\pm$2.227&0.824$\pm$0.080&6.006$\pm$0.638&4.744$\pm$0.494 \\ 
		\textbf{BDPN} & 28.973$\pm$1.714&0.824$\pm$0.093&7.089$\pm$0.864&4.856$\pm$0.570 \\ 
		\textbf{FusionNet} & 28.735$\pm$2.460&0.796$\pm$0.074&6.426$\pm$0.860&5.136$\pm$0.515 \\   
		\textbf{MUCNN} & 27.839$\pm$2.328&0.777$\pm$0.088&7.504$\pm$0.539&5.517$\pm$0.299 \\   
		\textbf{LAGNet} & 28.050$\pm$2.239&0.805$\pm$0.084&6.955$\pm$0.474&5.326$\pm$0.318 \\  
		\textbf{PMACNet} & 19.160$\pm$4.512&0.509$\pm$0.128&15.95$\pm$3.329&15.69$\pm$3.307 \\  	  
		\textbf{U2Net} & \textbf{\textcolor{red}{30.740}}$\pm$2.173 &\textbf{\textcolor{red}{0.849}}$\pm$0.085 & \textbf{\textcolor{red}{5.250}}$\pm$0.545 & \textbf{\textcolor{red}{4.070}}$\pm$0.392 \\ 
		\bottomrule
	\end{tabular}
	\label{testwv2}	
\end{table}

\subsection{Generalization}
\label{s43}
Generalization ability is a crucial concern for DL-based methods in the pansharpening task. If there is a significant difference between the testing and training datasets, some approaches may not perform well. We use 20 reduced-resolution samples acquired by WV2 to test all DL-based models trained on the WV3 dataset. The quantitative evaluation outcomes are presented in Tab.~\ref{testwv2}, and the U2Net yields the best results on all four metrics, indicating the strong generalization capability of our method. Notably, the inflexible and unreasonable structure of PMACNet \cite{9764690} significantly restricts its generalization ability, leading to extremely unsatisfactory outcomes.

\subsection{Comparison of Parameter Numbers}
\begin{figure}[t]
	\begin{center}
		\begin{minipage}[t]{0.95\linewidth}
			\begin{minipage}[t]{0.495\linewidth}
				{\includegraphics[width=1\linewidth]{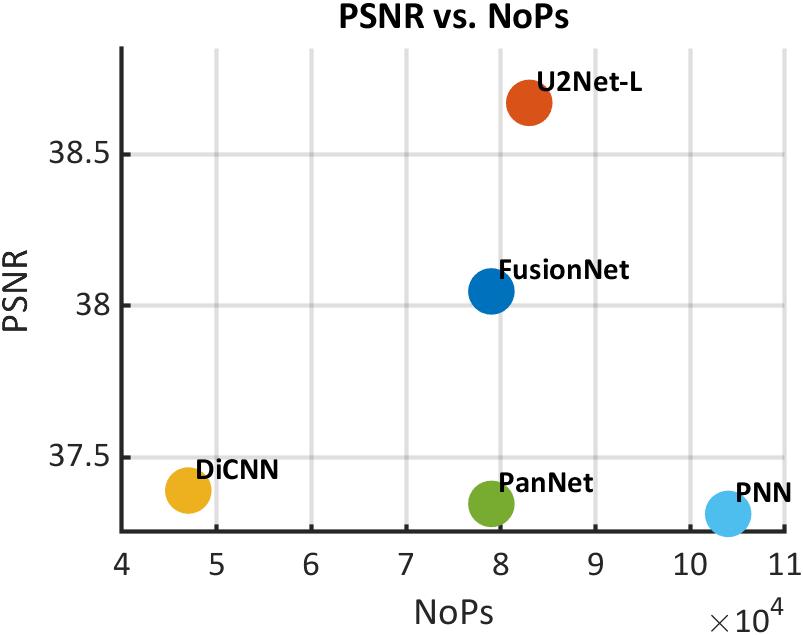}}
				{\includegraphics[width=1\linewidth]{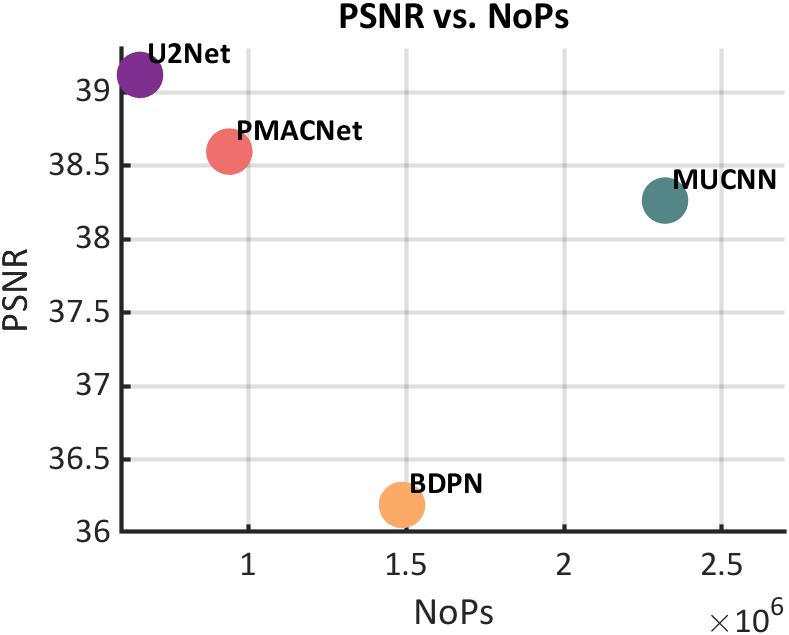}}
				
			\end{minipage}
			\begin{minipage}[t]{0.495\linewidth}
				{\includegraphics[width=1\linewidth]{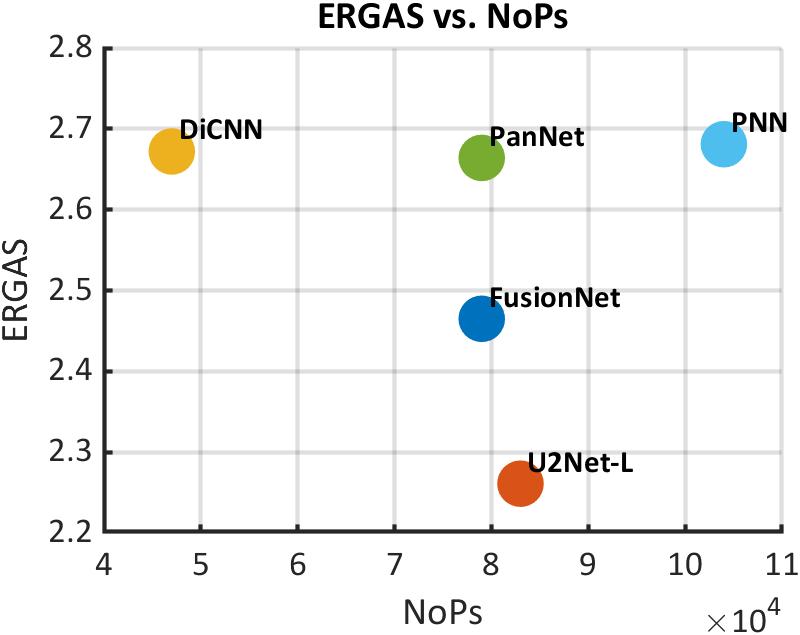}}
				{\includegraphics[width=1\linewidth]{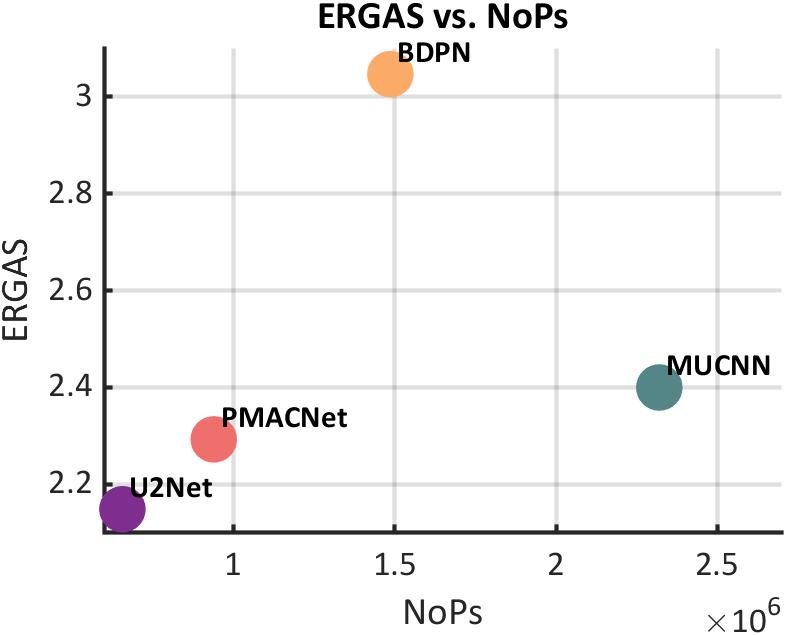}}
			\end{minipage}
		\end{minipage}
	\end{center}
	\caption{The comparisons of NoPs. The first row: comparisons of lightweight networks ($\leq1\times10^5$ parameters) based on PSNR and ERGAS indexes. The second row: comparisons of heavyweight networks ($\geq5\times10^5$ parameters) based on the same quality indexes.\label{abb}}
\end{figure}

We categorize the DL-based pansharpening methods into two groups based on their number of parameters (NoPs). Specifically, models with less than $1\times10^5$ parameters are considered lightweight networks, whereas those with more than $5\times10^5$ parameters are classified as heavyweight networks. The U2Net is a heavyweight network, which prompts us to develop a lightweight version called U2Net-L to demonstrate the superiority of our method more effectively. To ensure fairness, we compare U2Net-L with lightweight networks and U2Net with heavyweight networks. Fig.~\ref{abb} shows the comparisons of NoPs on 20 reduced-resolution samples acquired by WV3. Both U2Net-L and U2Net achieve exceptional performance within their respective categories, demonstrating the superiority of our framework. For more details, kindly refer to the \emph{Sup. Mat}.

\begin{table}[t]
	\centering\renewcommand\arraystretch{1.4}\setlength{\tabcolsep}{5.8pt}
	\footnotesize
	\caption{Ablation study on 20 reduced-resolution samples acquired by WV2. (\textcolor{red}{Red}: best; \textcolor{blue}{Blue}: second best).}
	\begin{tabular}{ccccc}
		\toprule
		\textbf{Method} & PSNR($\pm$std)& Q8($\pm$std) & SAM($\pm$std) & ERGAS($\pm$std)\\  
		\midrule
		\textbf{V1} & 29.849$\pm$2.171&0.830$\pm$0.087&5.773$\pm$0.731&4.512$\pm$0.740 \\   
		\textbf{V2} & 30.295$\pm$2.324&0.839$\pm$0.083&5.520$\pm$0.634&4.281$\pm$0.380 \\      
		\textbf{V3} & \textcolor{blue}{30.394}$\pm$2.380&0.841$\pm$0.081&\textbf{\textcolor{red}{5.165}}$\pm$0.610&\textcolor{blue}{4.248}$\pm$0.376 \\  
		\textbf{V4} & 30.104$\pm$2.246&\textcolor{blue}{0.848}$\pm$0.086&5.575$\pm$0.691&4.380$\pm$0.535 \\  
		\textbf{U2Net} & \textbf{\textcolor{red}{30.740}}$\pm$2.173 &\textbf{\textcolor{red}{0.849}}$\pm$0.085 & \textcolor{blue}{5.250}$\pm$0.545 & \textbf{\textcolor{red}{4.070}}$\pm$0.392 \\ 
		\bottomrule
	\end{tabular}
	\label{abl}
\end{table}

\subsection{Ablation Study}
To validate the effectiveness of our method, we create four variants of the U2Net. In the first variant (V1), we employ a single-branch U-shape network to extract spatial and spectral features uniformly while maintaining the original structure of the S2Block. The purpose of V1 is to demonstrate that the double-branch network is more effective in capturing diverse information compared to the single-branch one. The second variant (V2) retains the double U-shape network architecture but replaces the S2Blocks with concatenation operations. The V2 is designed to confirm the superiority of S2Blocks in information integration. In the third variant (V3), the S2Block only produces spatial self-correlation matrices and combines them with the spectral feature map. As for the fourth variant (V4), only spectral self-correlation matrices are generated and merged with the spatial feature map. 

We perform experiments on 20 reduced-resolution samples acquired by WV2. The results are presented in Tab.~\ref{abl}, and the U2Net yields the best overall performance, proving the effectiveness of our method. Further explanation and discussion on the ablation study can be found in the \emph{Sup. Mat}. \label{ablations}

\begin{figure*}[t]
	\begin{center}
		\begin{minipage}[t]{0.96\linewidth}
			\begin{minipage}[t]{0.096\linewidth}
				{\includegraphics[width=1\linewidth]{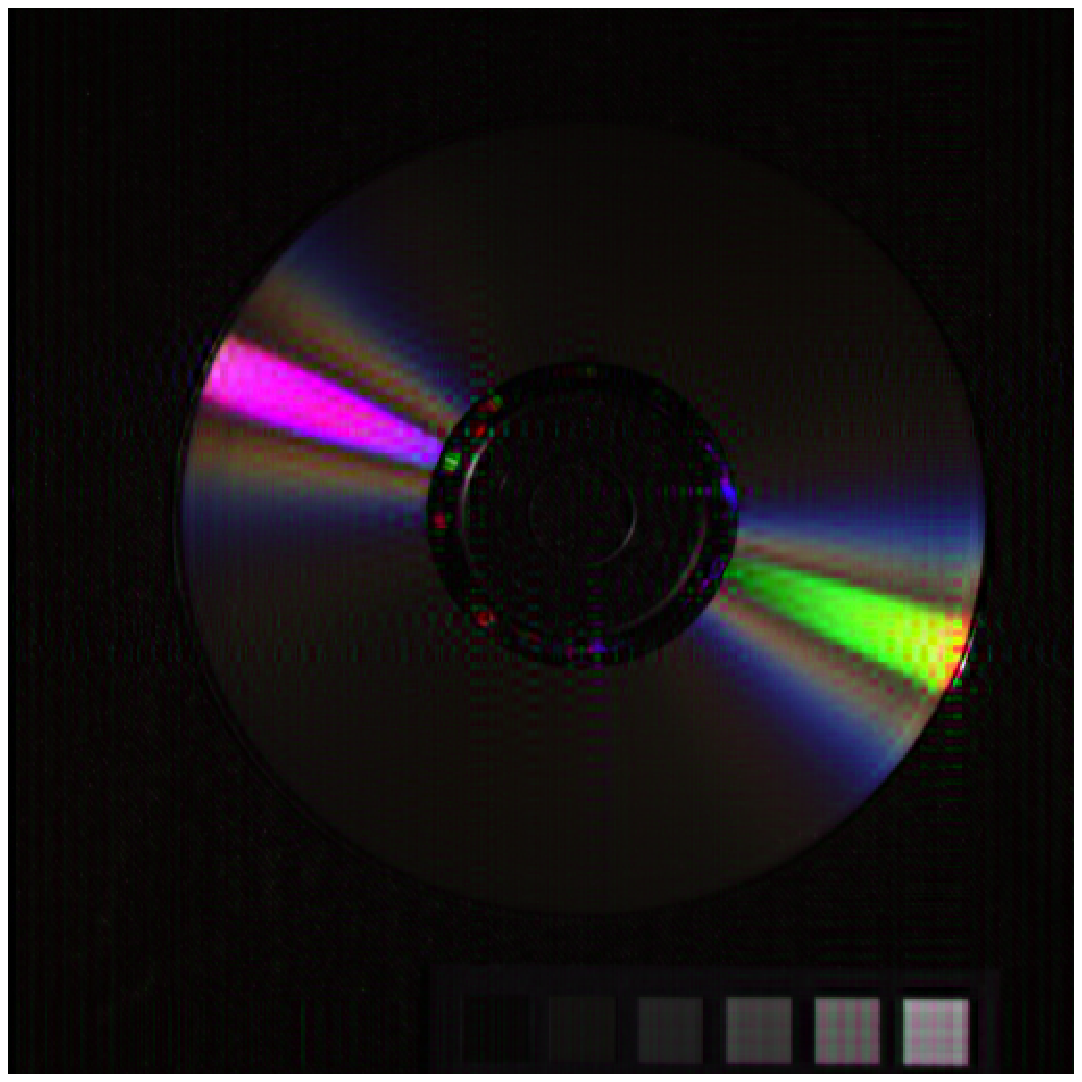}}
				{\includegraphics[width=1\linewidth]{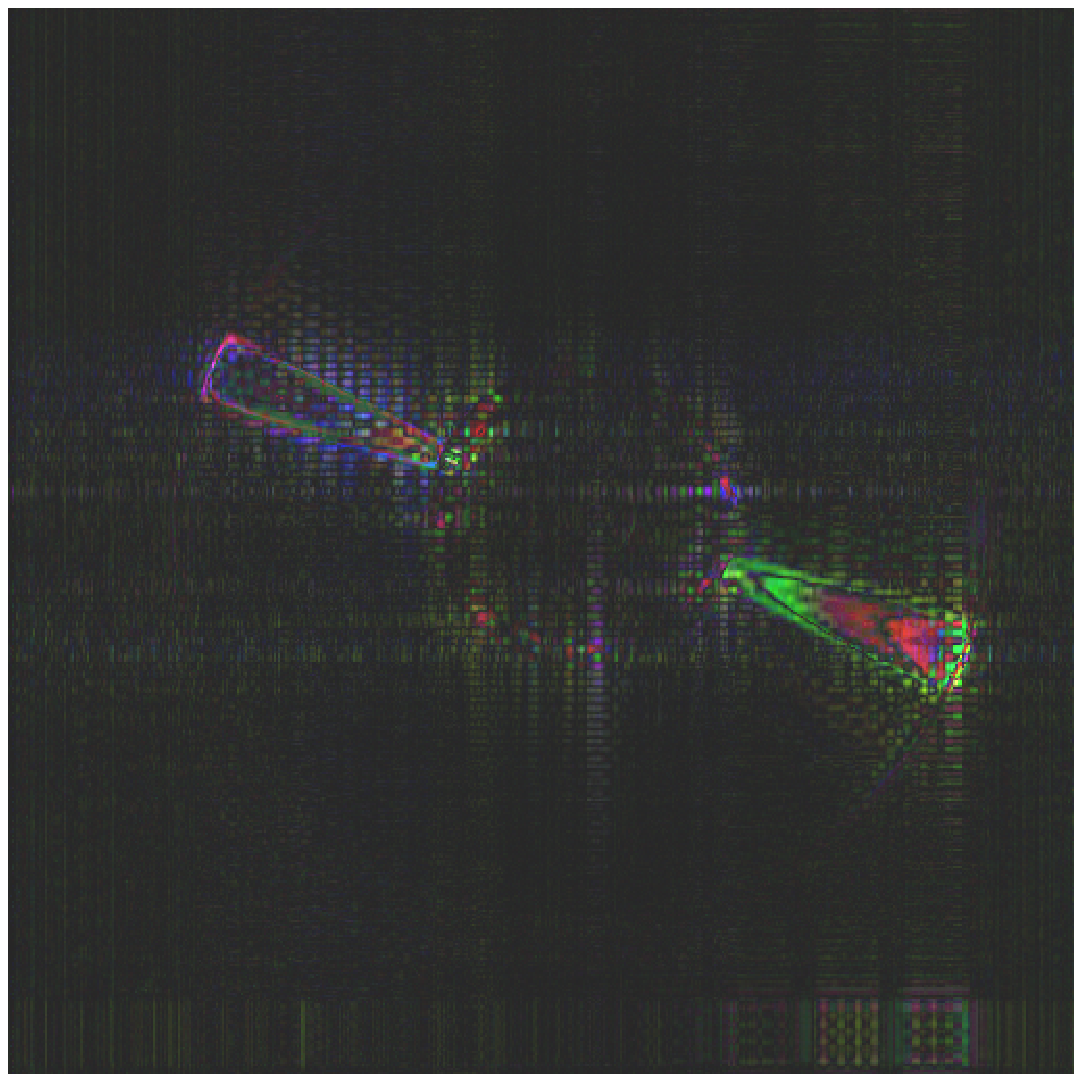}}
				{CSTF}
				\centering
				
			\end{minipage}
			\begin{minipage}[t]{0.096\linewidth}
				{\includegraphics[width=1\linewidth]
					{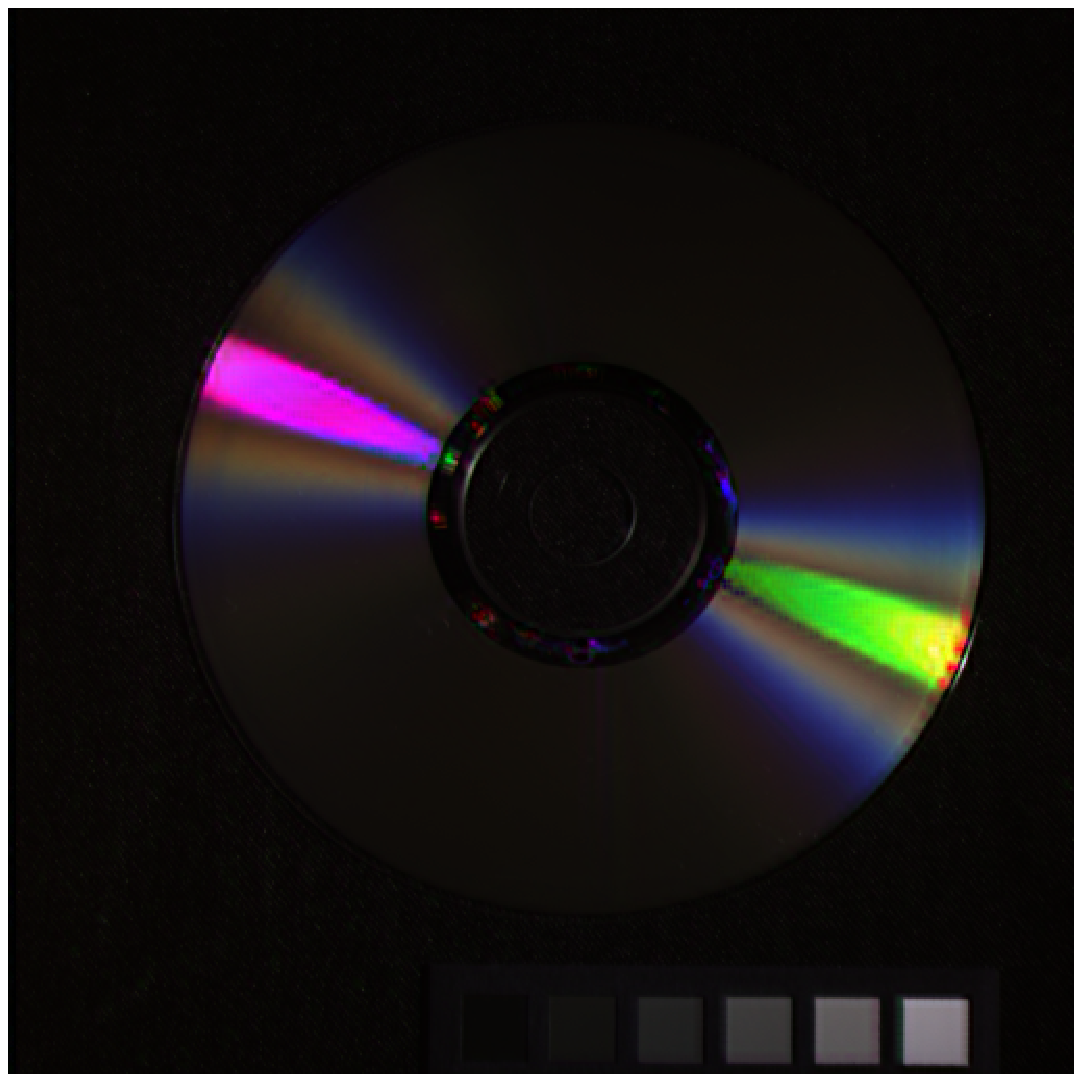}}
				{\includegraphics[width=1\linewidth]{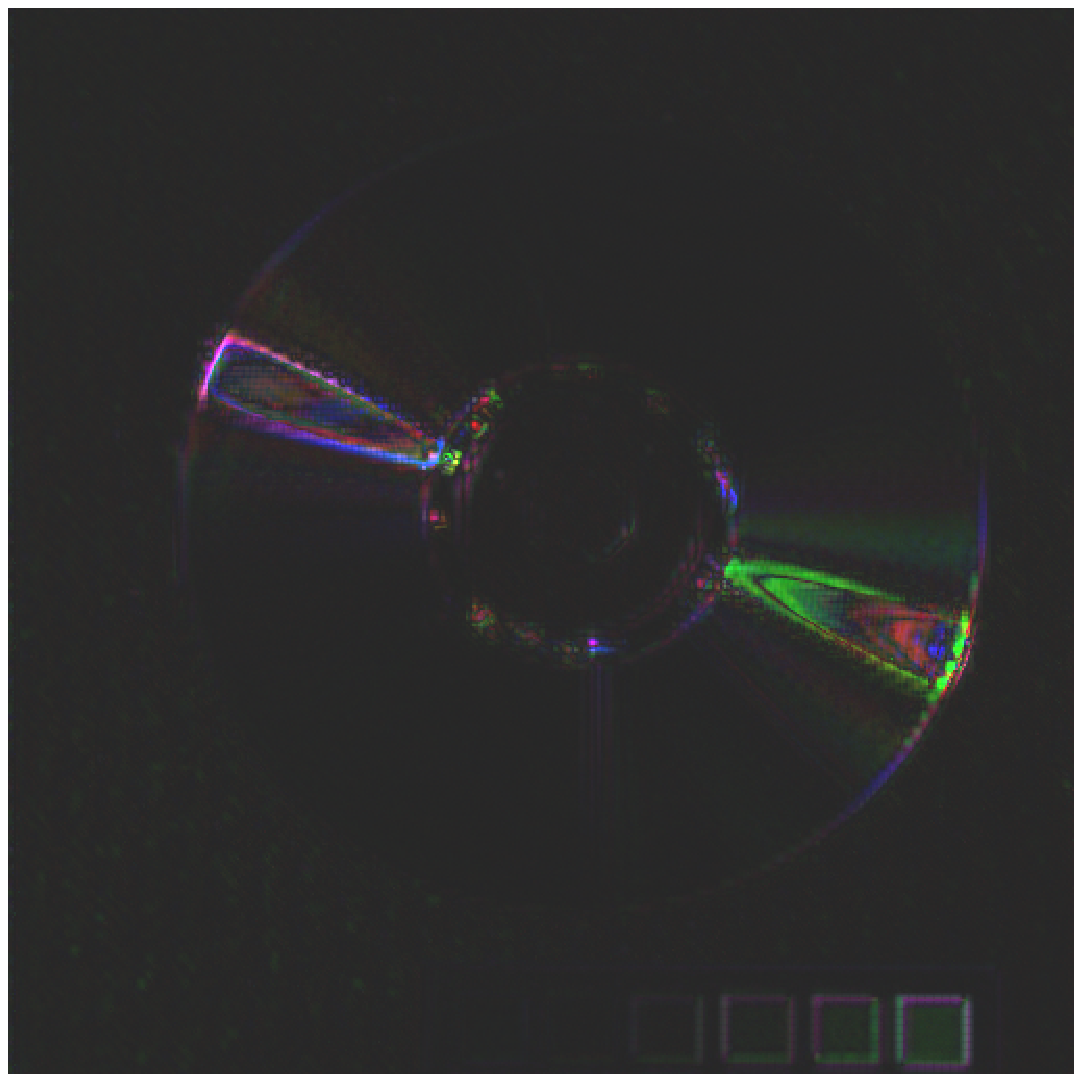}}
				{LTMR}
				\centering
				
			\end{minipage}
			\begin{minipage}[t]{0.096\linewidth}
				{\includegraphics[width=1\linewidth]{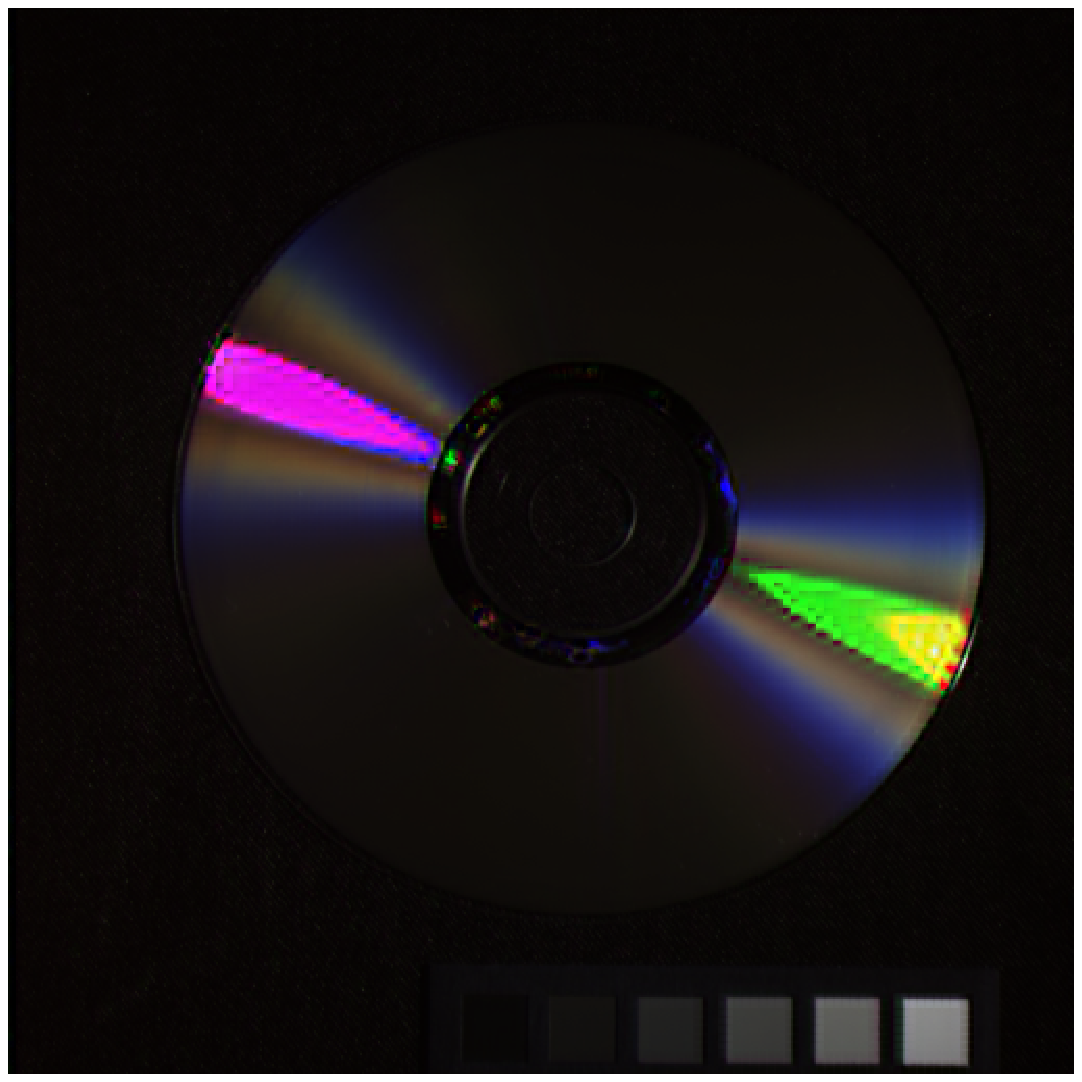}}
				{\includegraphics[width=1\linewidth]{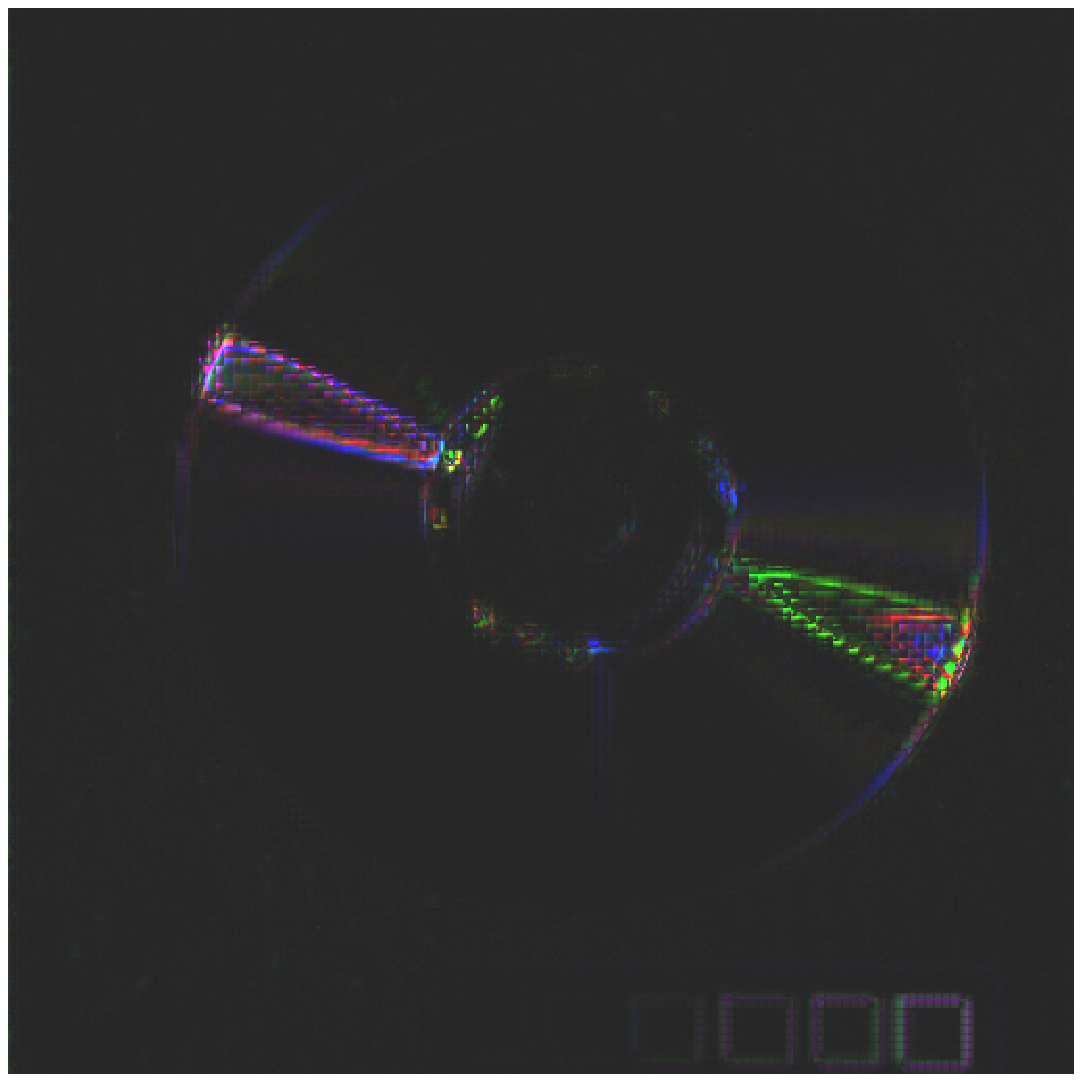}}
				{LTTR}
				\centering
				
			\end{minipage}
			\begin{minipage}[t]{0.096\linewidth}
				{\includegraphics[width=1\linewidth]{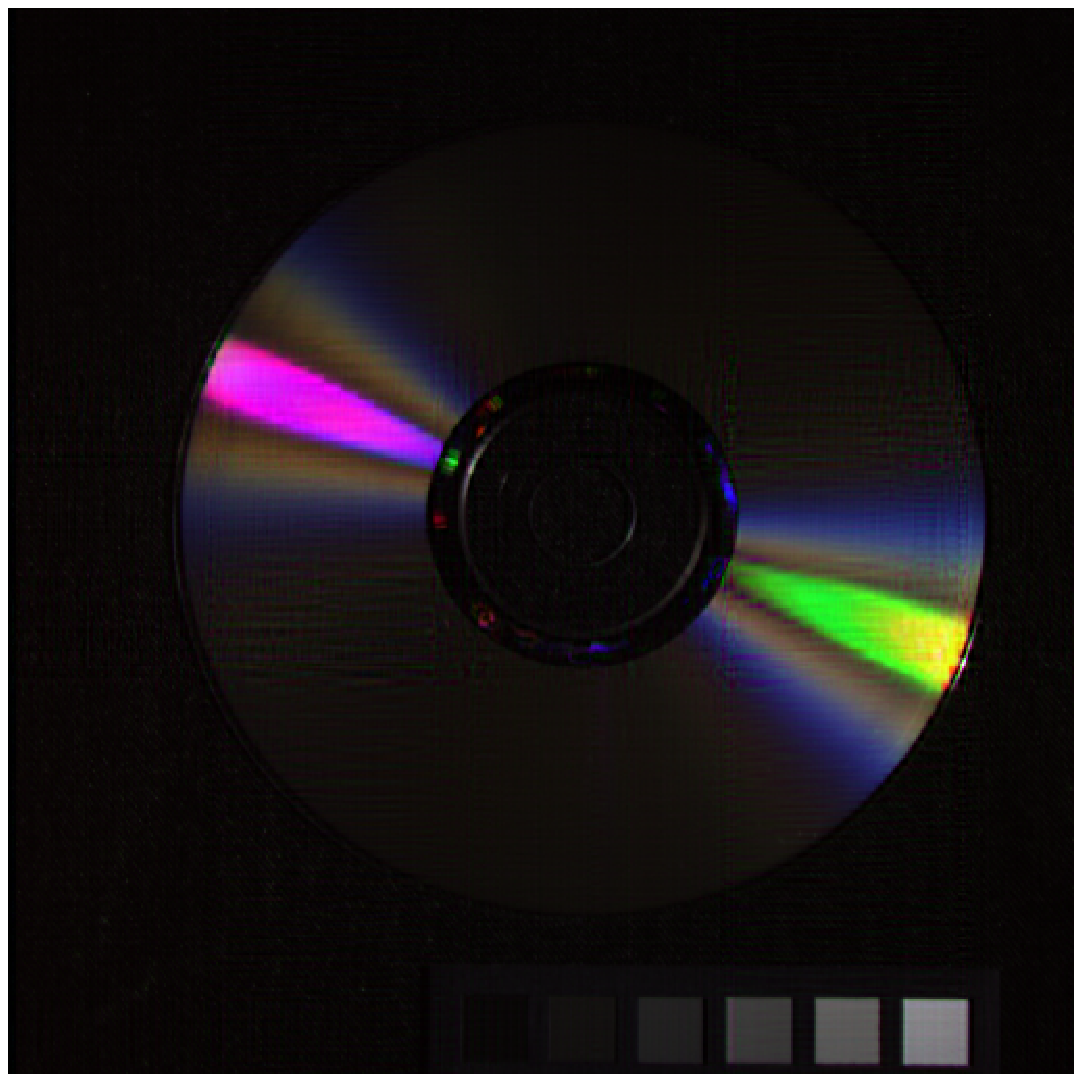}}
				{\includegraphics[width=1\linewidth]{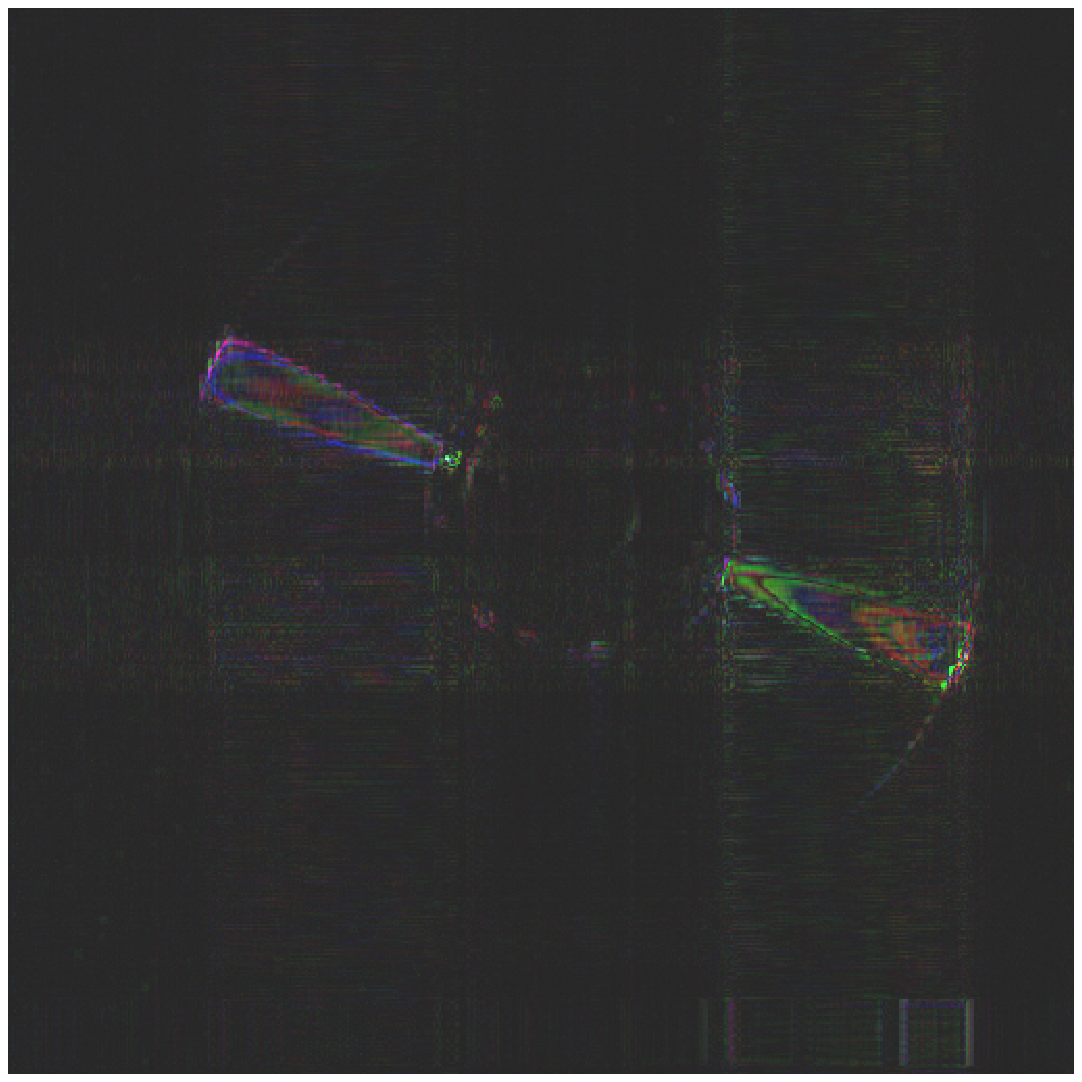}}
				{UTV}
				\centering
				
			\end{minipage}
			\begin{minipage}[t]{0.096\linewidth}
				{\includegraphics[width=1\linewidth]{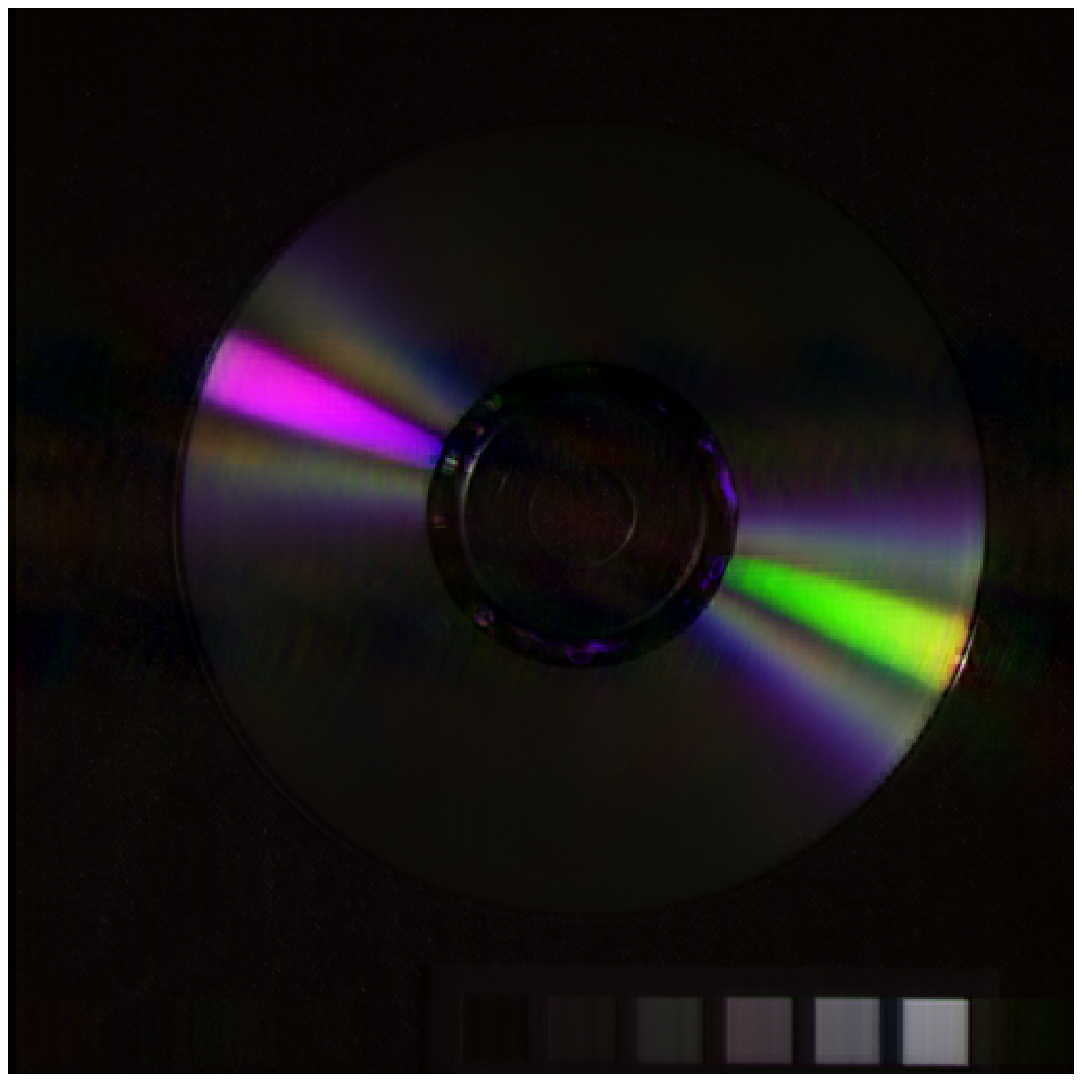}}
				{\includegraphics[width=1\linewidth]{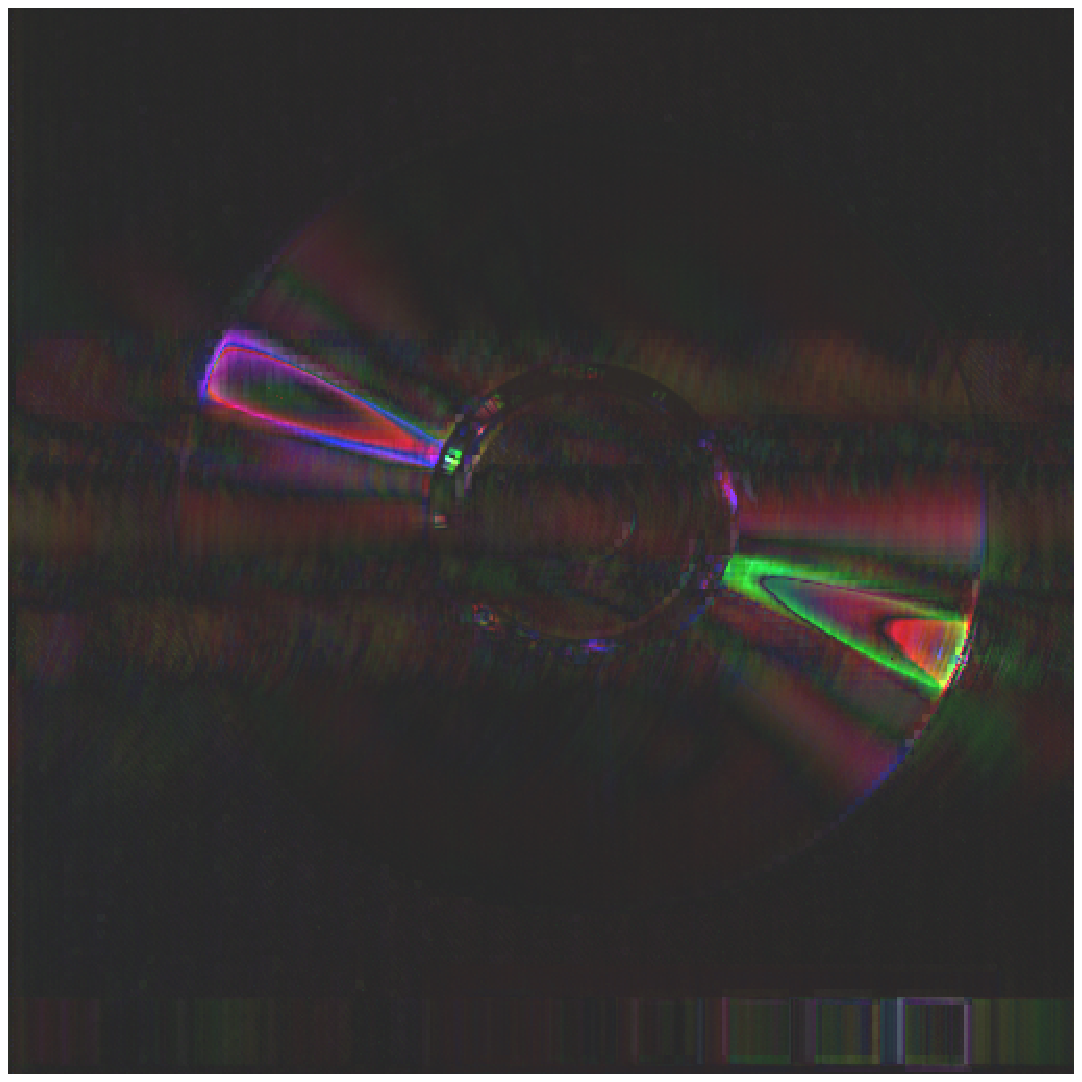}}
				{IR-TenSR}
				\centering
				
			\end{minipage}
			\begin{minipage}[t]{0.096\linewidth}
				{\includegraphics[width=1\linewidth]{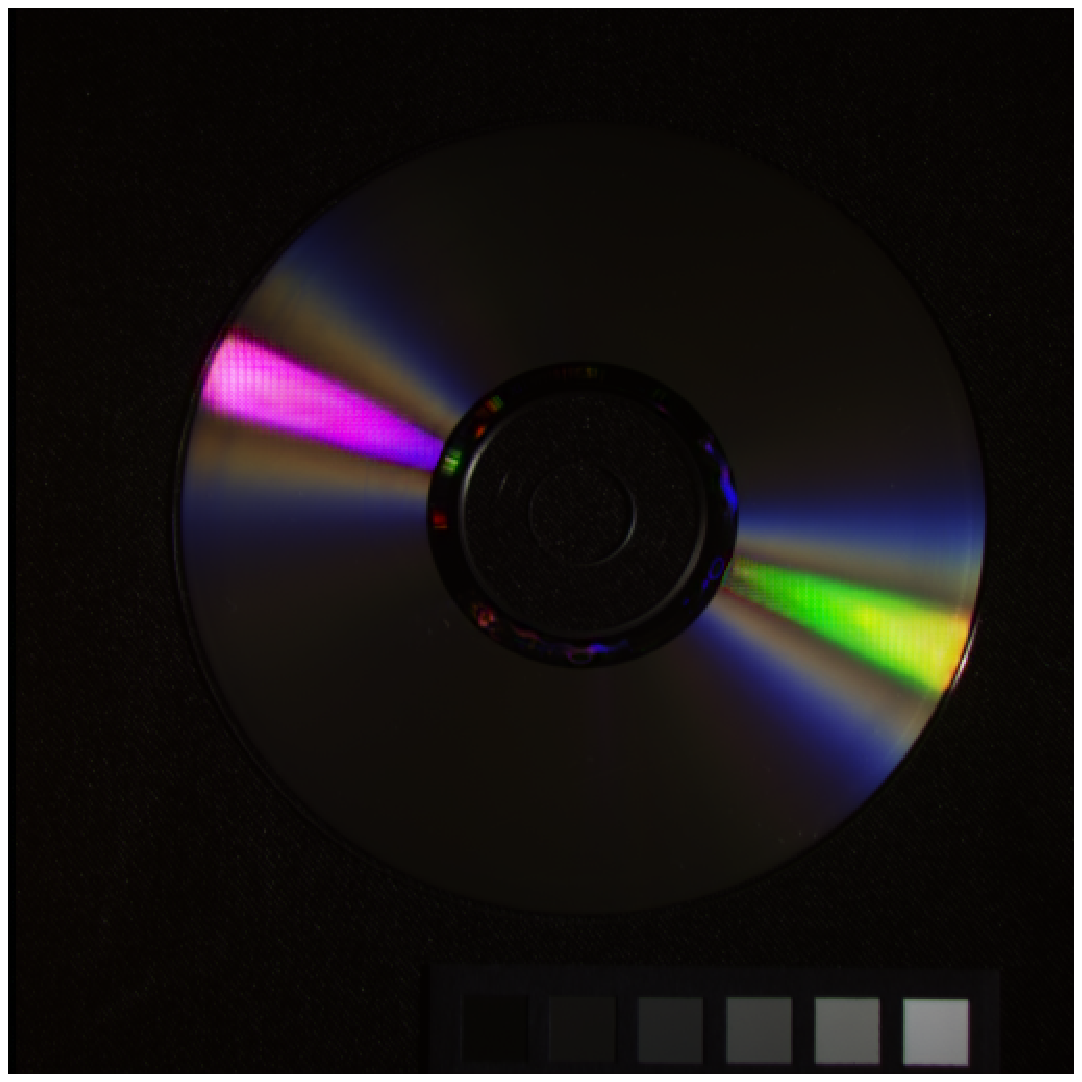}}
				{\includegraphics[width=1\linewidth]{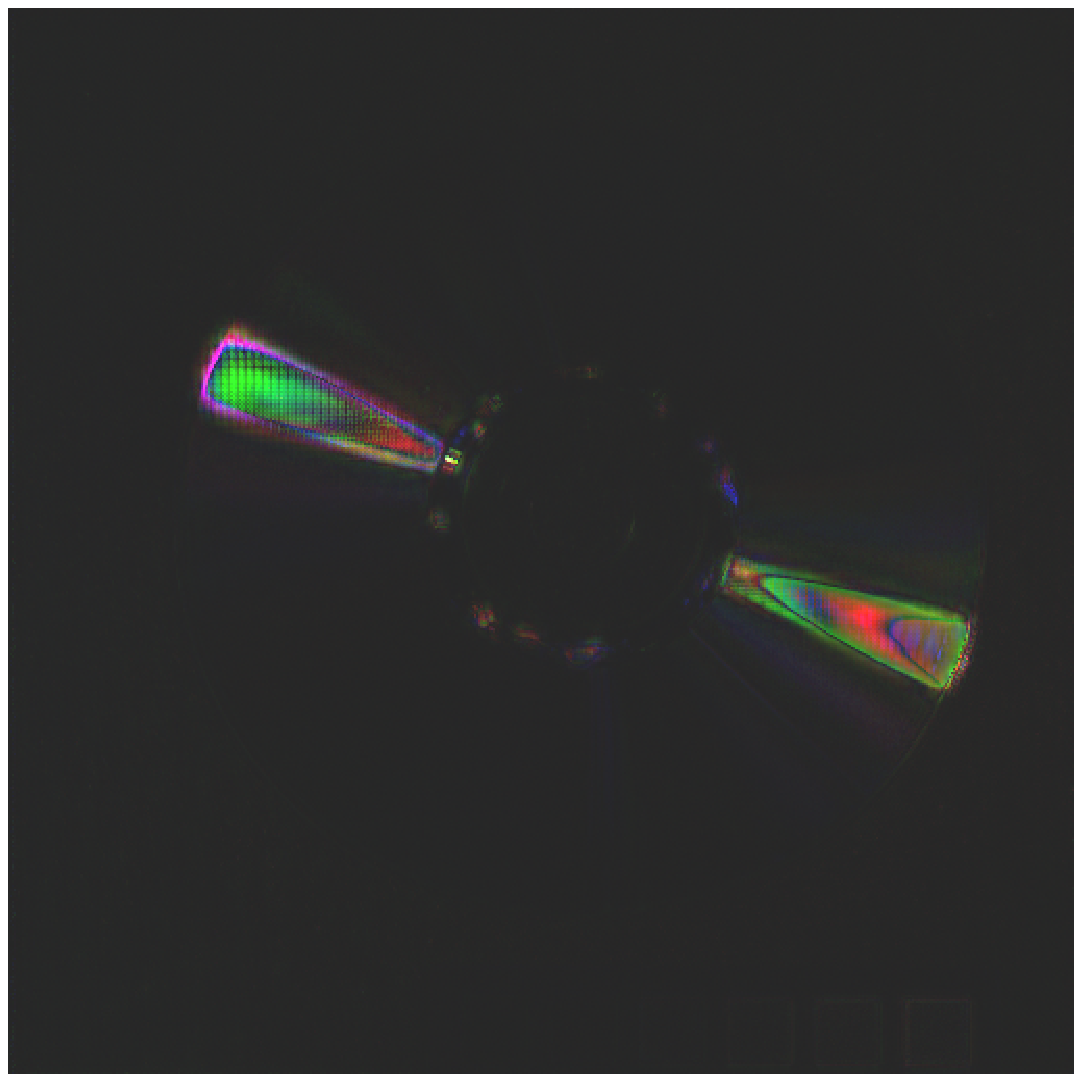}}
				{ResTFNet}
				\centering
				
			\end{minipage}
			\begin{minipage}[t]{0.096\linewidth}
				{\includegraphics[width=1\linewidth]{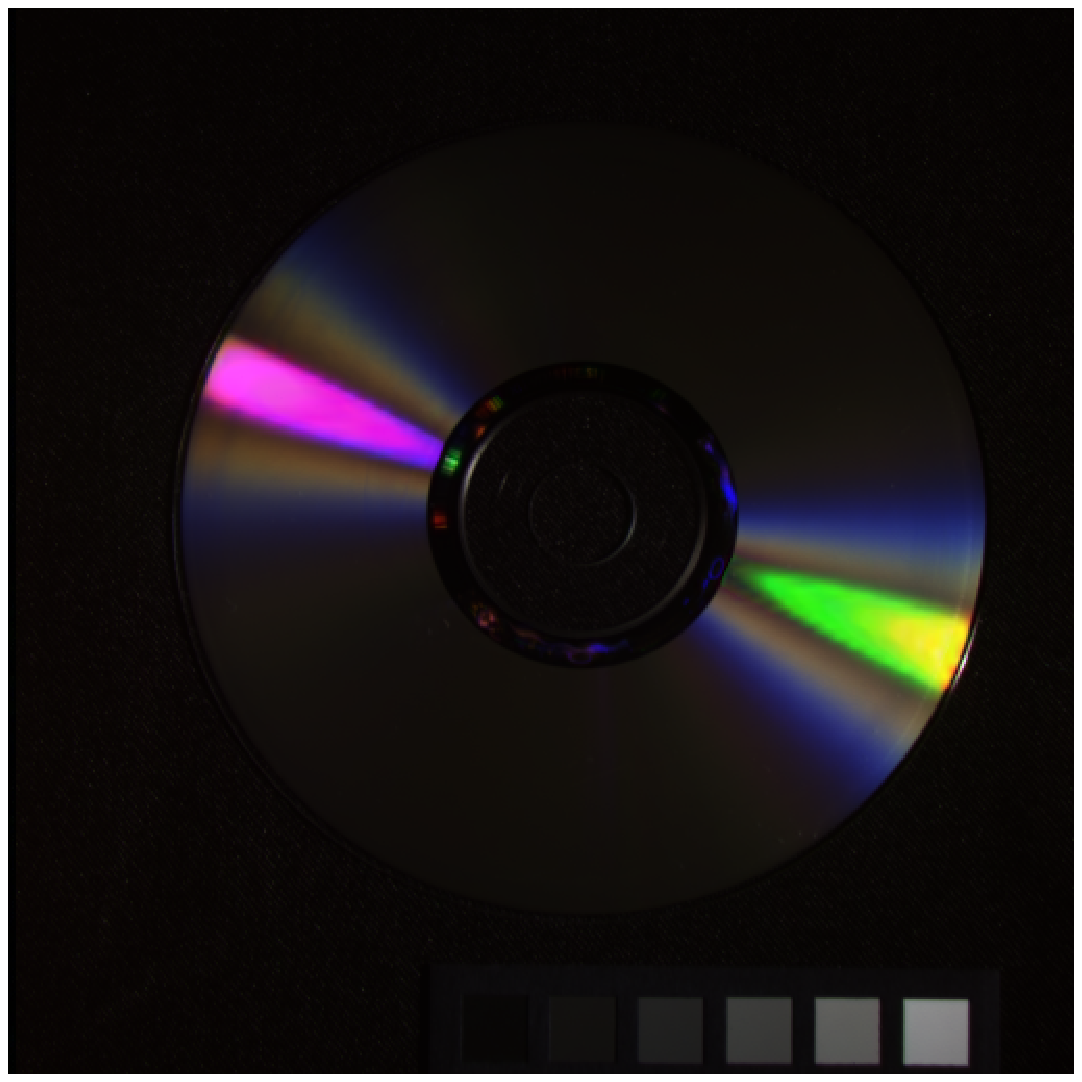}}
				{\includegraphics[width=1\linewidth]{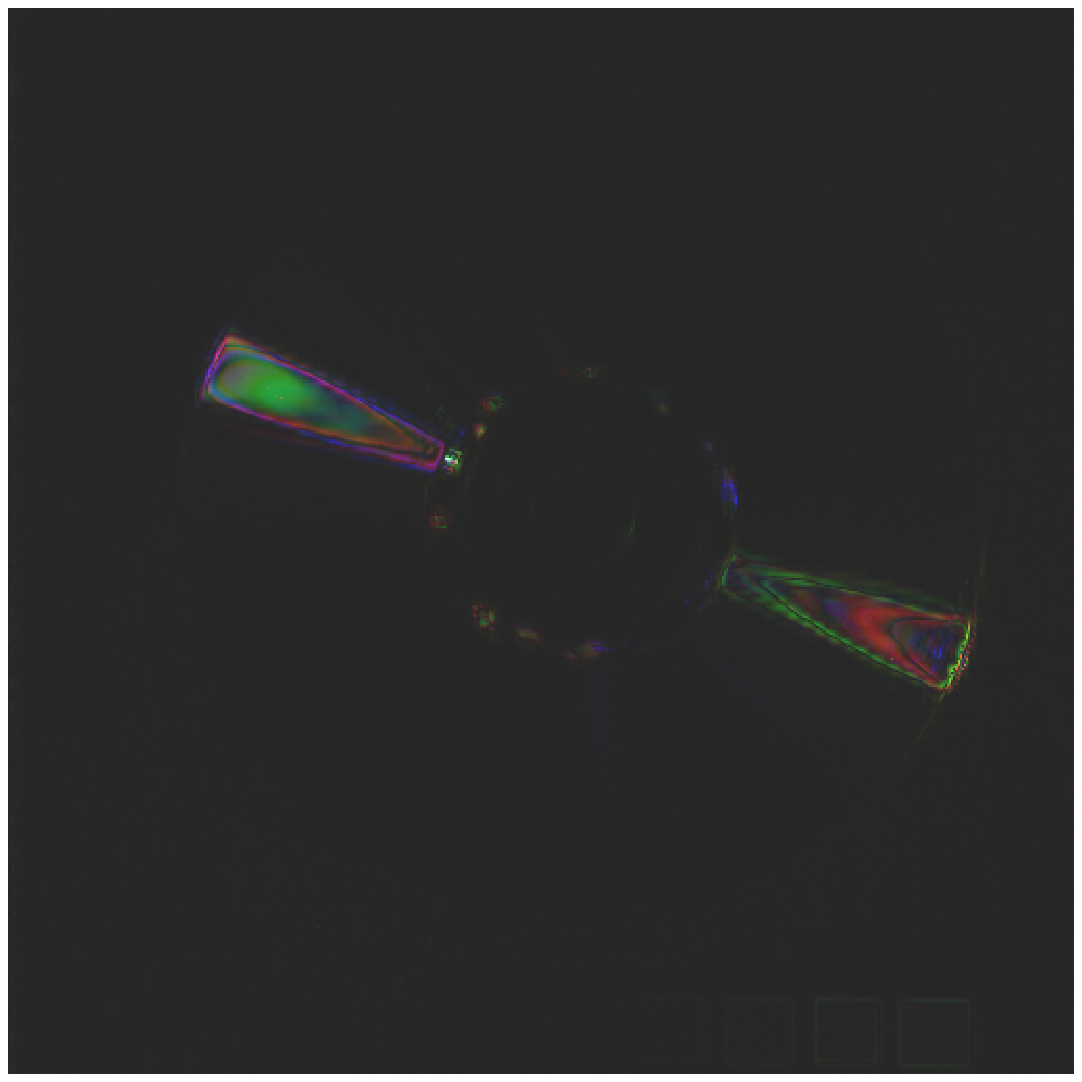}}
				{SSRNet}
				\centering
				
			\end{minipage}
			\begin{minipage}[t]{0.096\linewidth}
				{\includegraphics[width=1\linewidth]{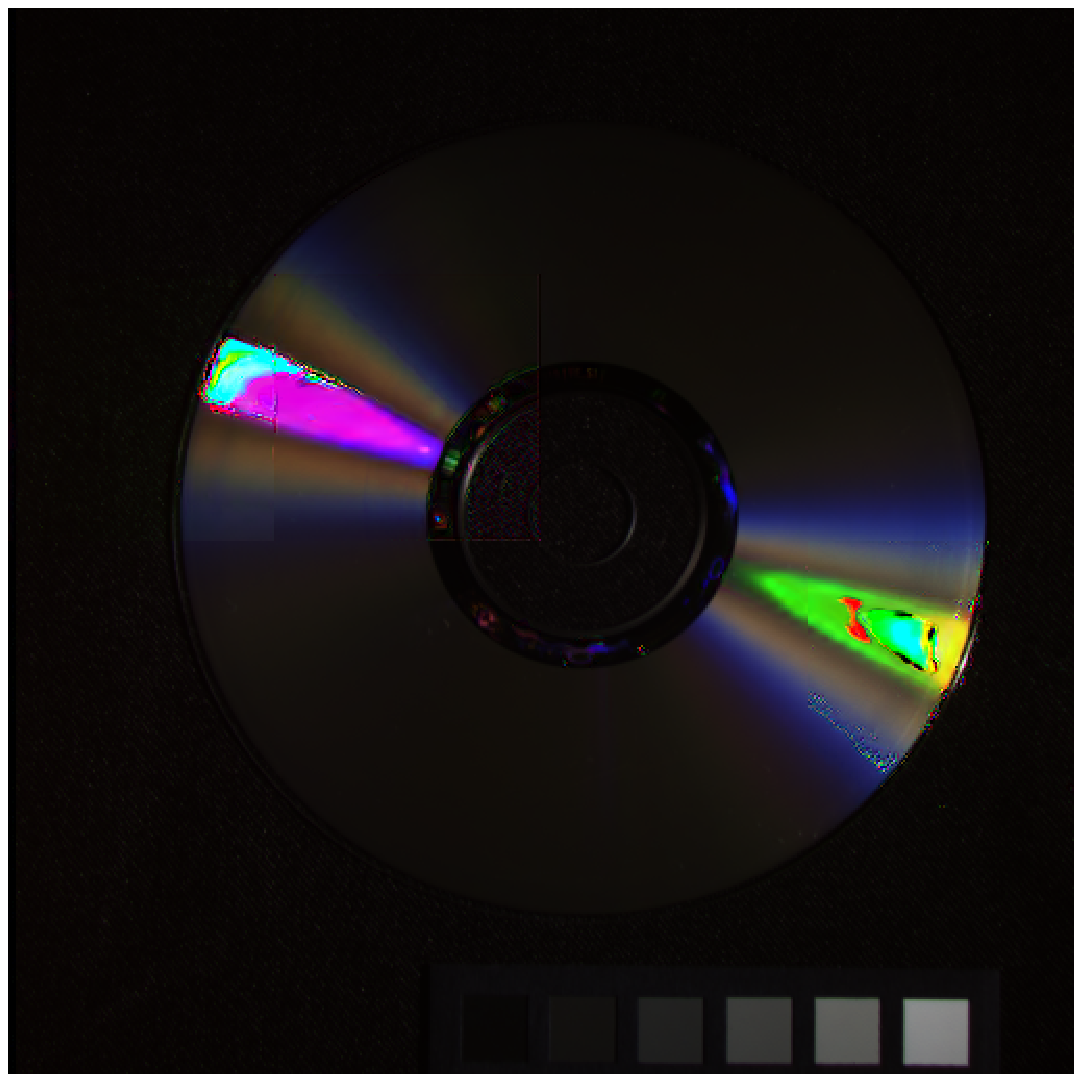}}
				{\includegraphics[width=1\linewidth]{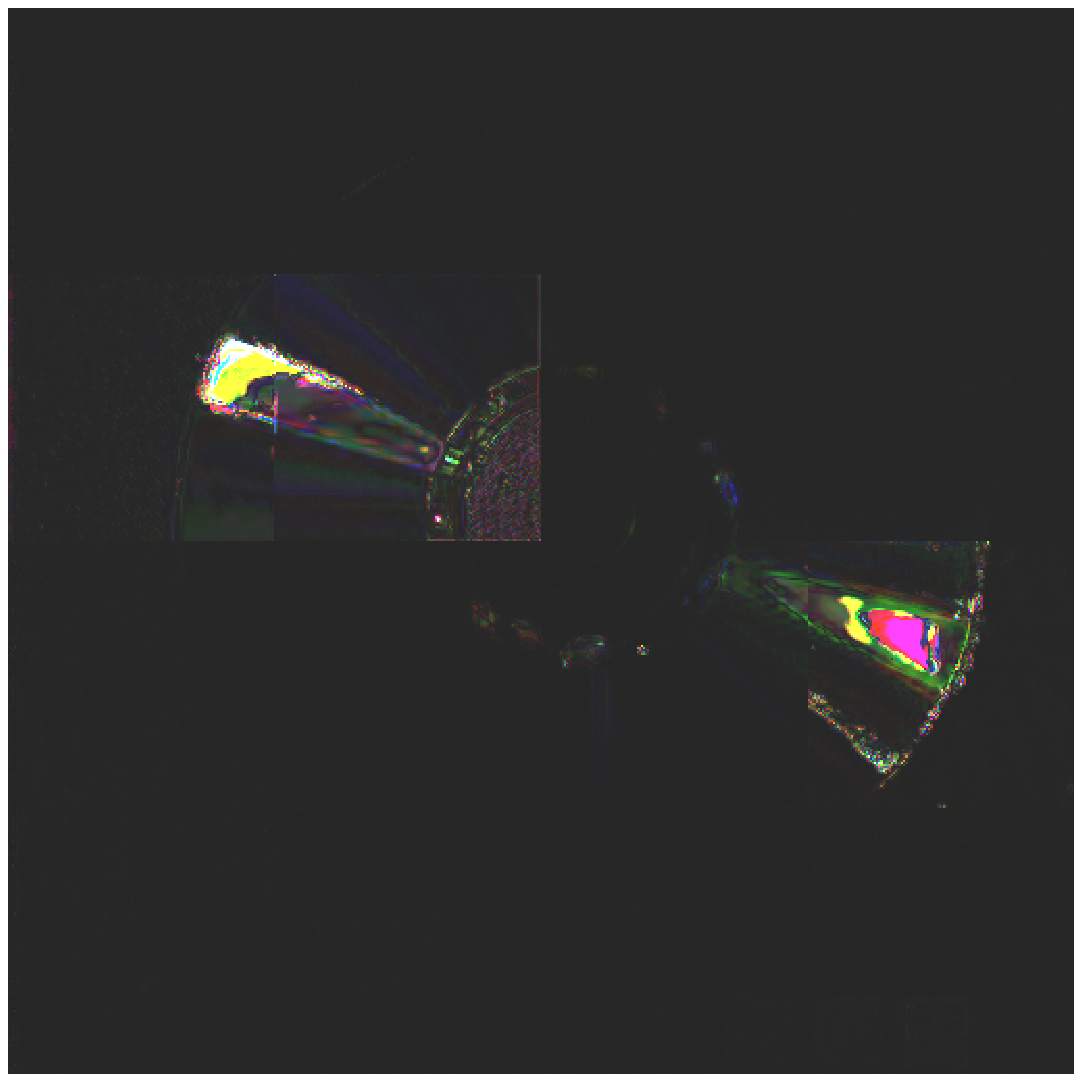}}
				{Fusformer}
				\centering
				
			\end{minipage}
			\begin{minipage}[t]{0.096\linewidth}
				{\includegraphics[width=1\linewidth]{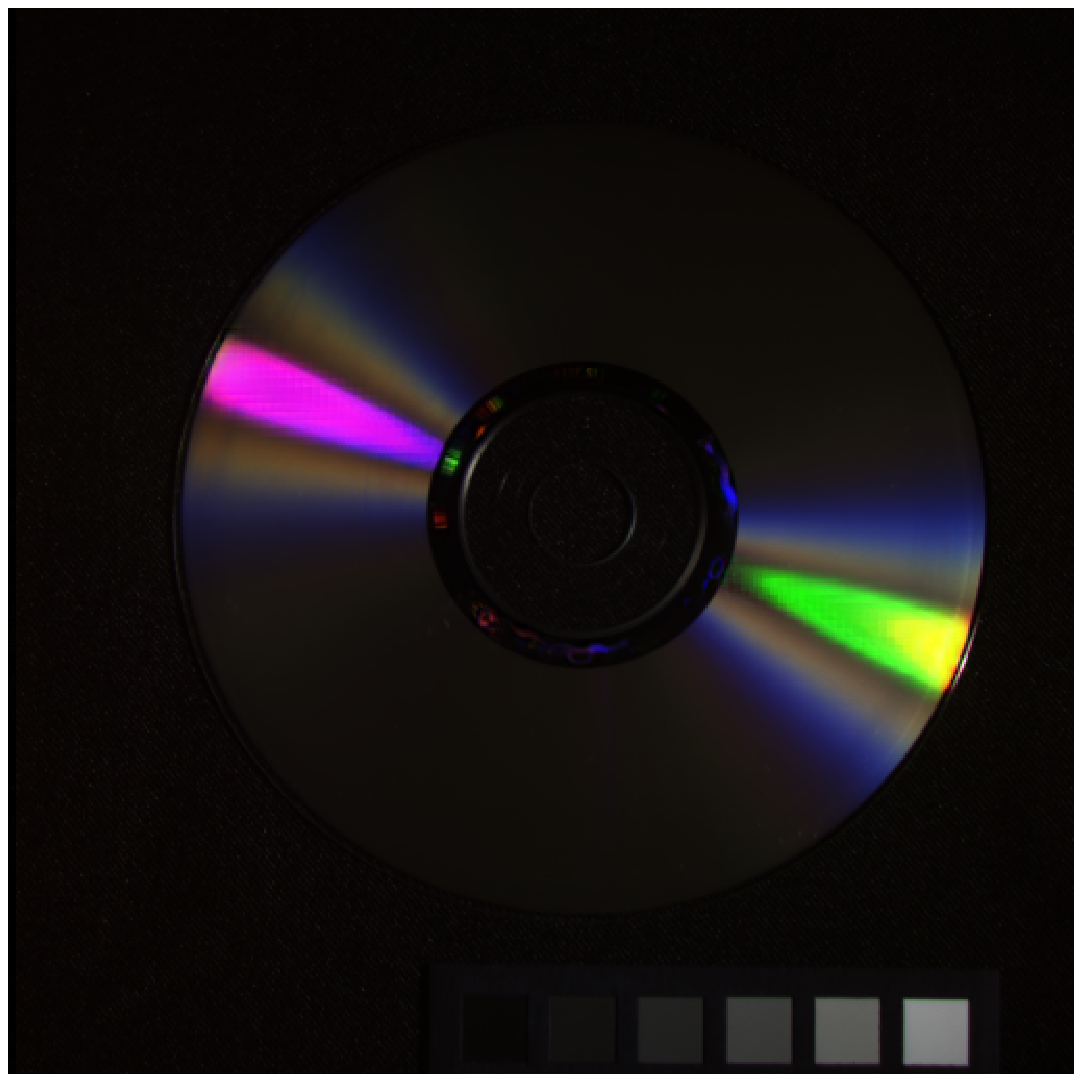}}
				{\includegraphics[width=1\linewidth]{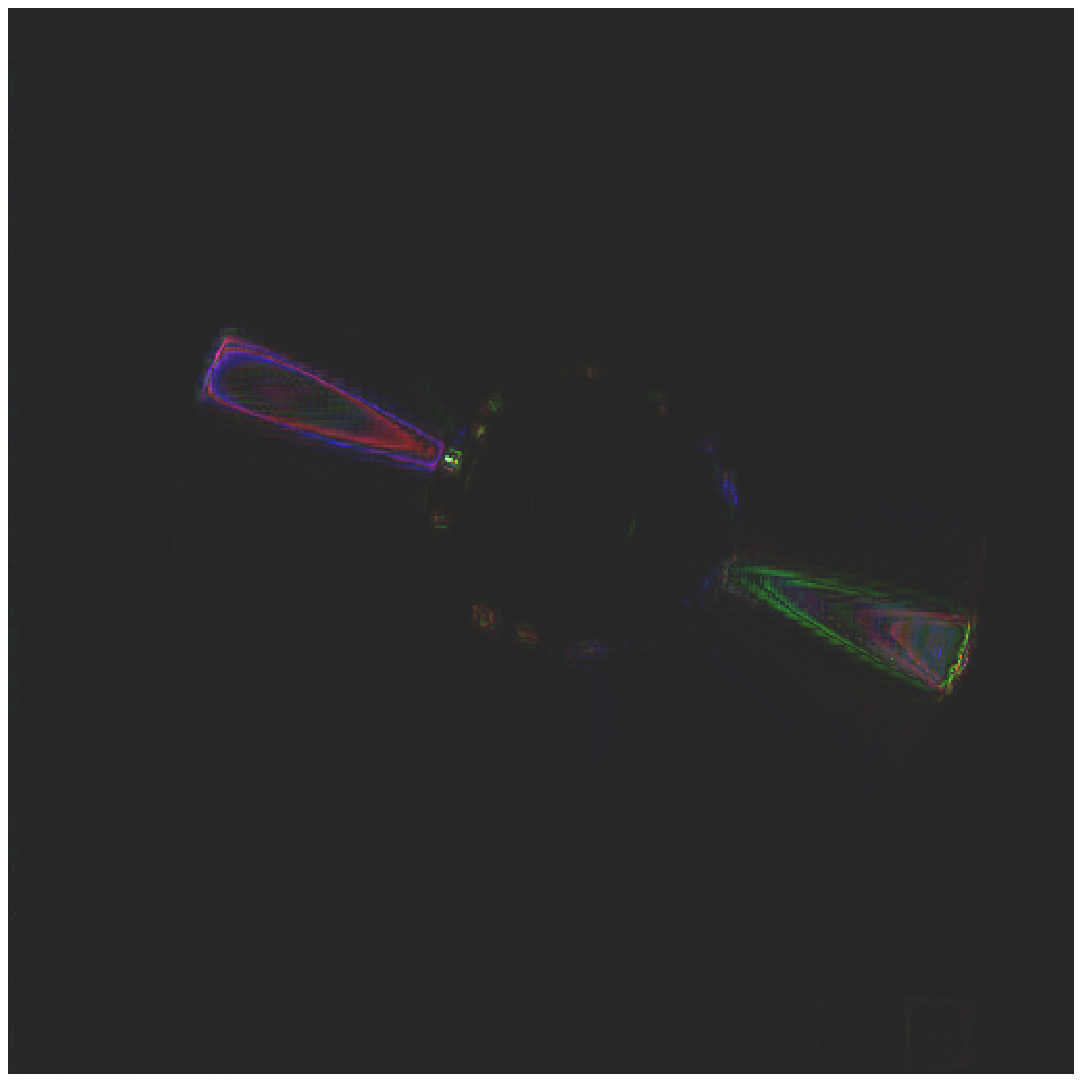}}
				{U2Net}
				\centering
				
			\end{minipage}
			\begin{minipage}[t]{0.096\linewidth}
				{\includegraphics[width=1\linewidth]{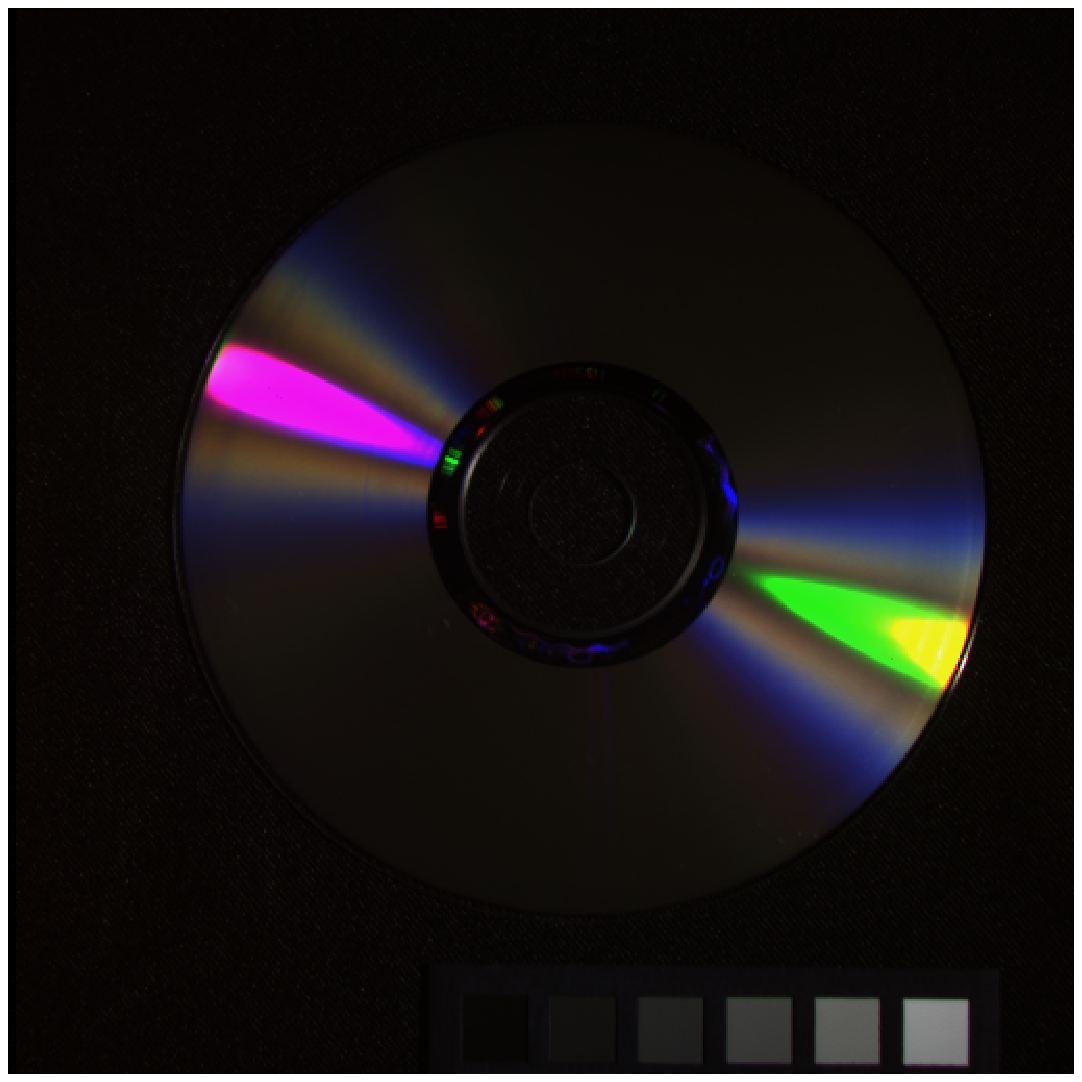}}
				{\includegraphics[width=1\linewidth]{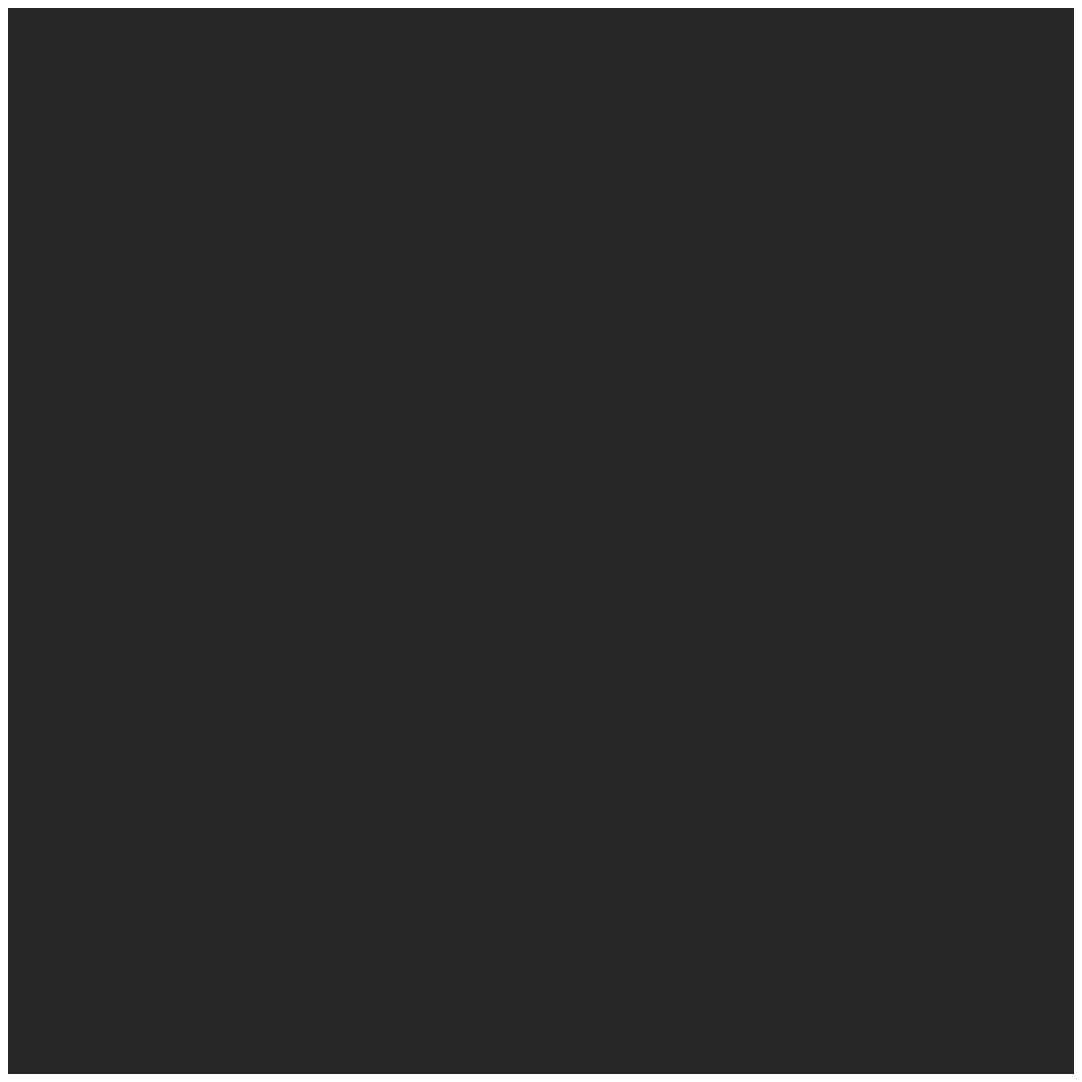}}
				{GT}
				\centering
				
			\end{minipage}
			\vspace{-7pt}
		\end{minipage}
	\end{center}
	\caption{Qualitative evaluation results on a CAVE testing sample. The first row: natural color maps. The second row: AEMs.\label{comparison2}}
\end{figure*}

\begin{figure*}[h!]
	\begin{center}
		\begin{minipage}[t]{0.98\linewidth}
			\begin{minipage}[t]{0.33\linewidth}
				{\includegraphics[width=1\linewidth]{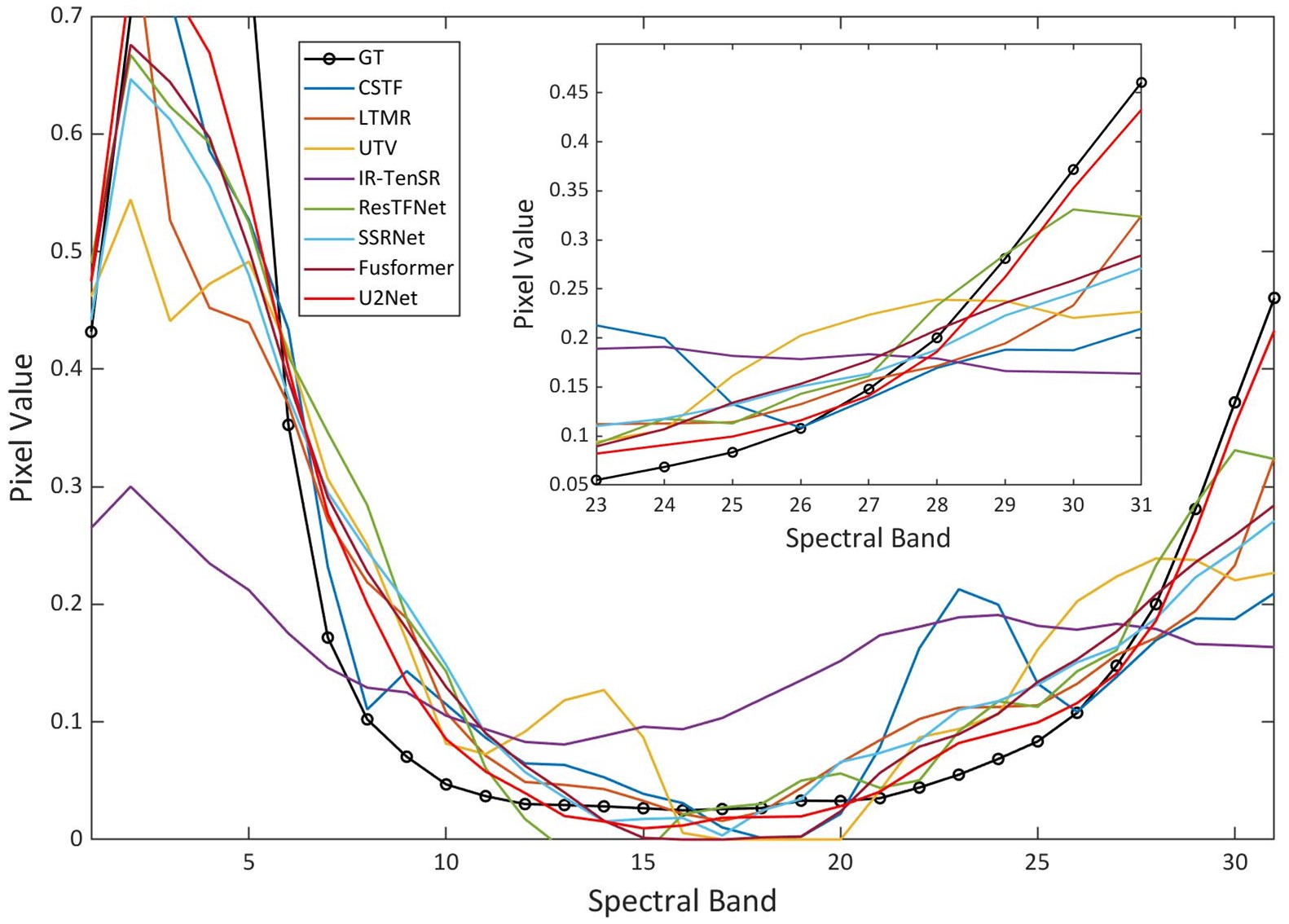}}
				{Spectral vectors at (293, 233)}
				\centering
				
			\end{minipage}
			\begin{minipage}[t]{0.33\linewidth}
				{\includegraphics[width=1\linewidth]{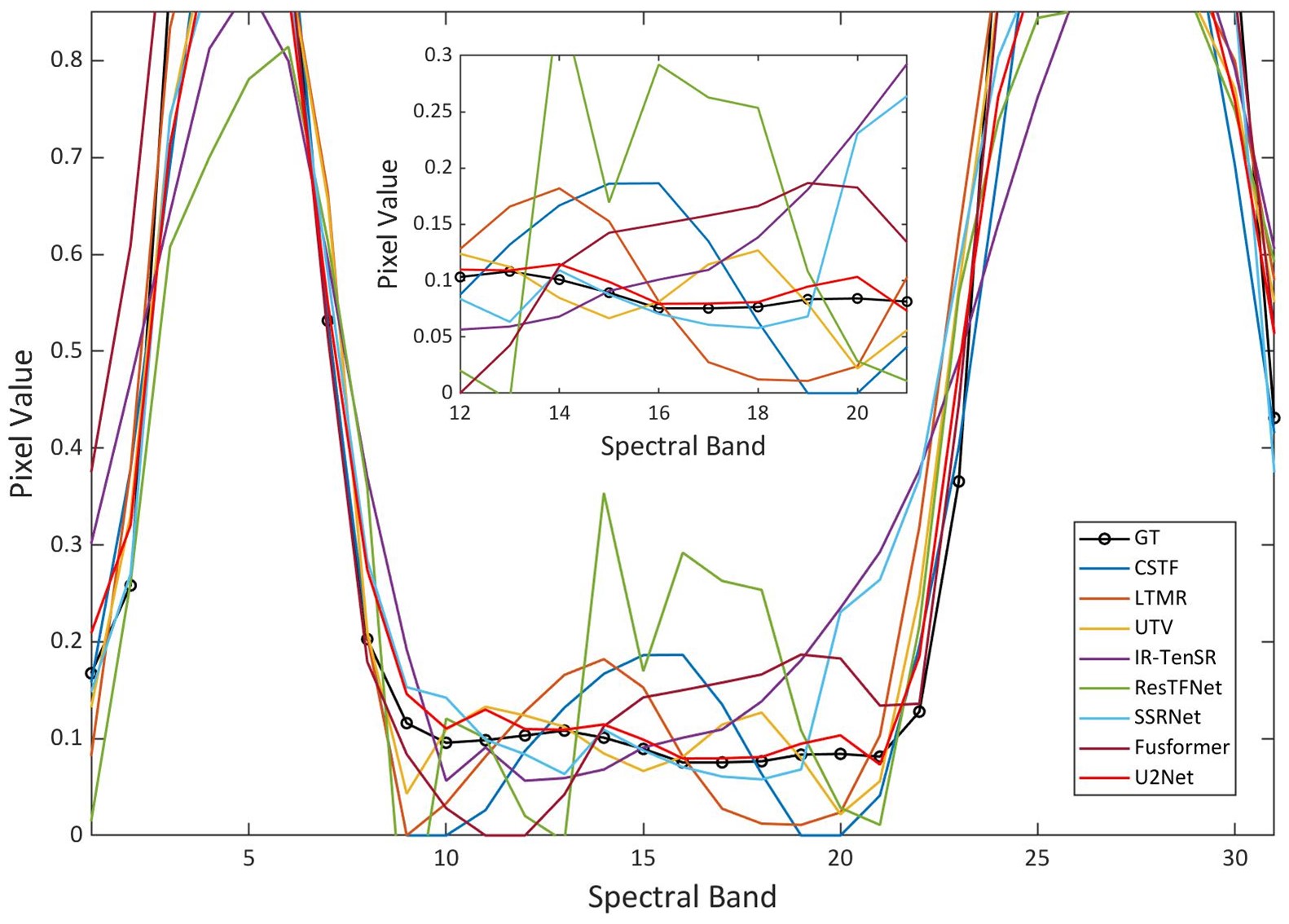}}
				{Spectral vectors at (194, 172)}
				\centering
				
			\end{minipage}
			\begin{minipage}[t]{0.33\linewidth}
				{\includegraphics[width=1\linewidth]{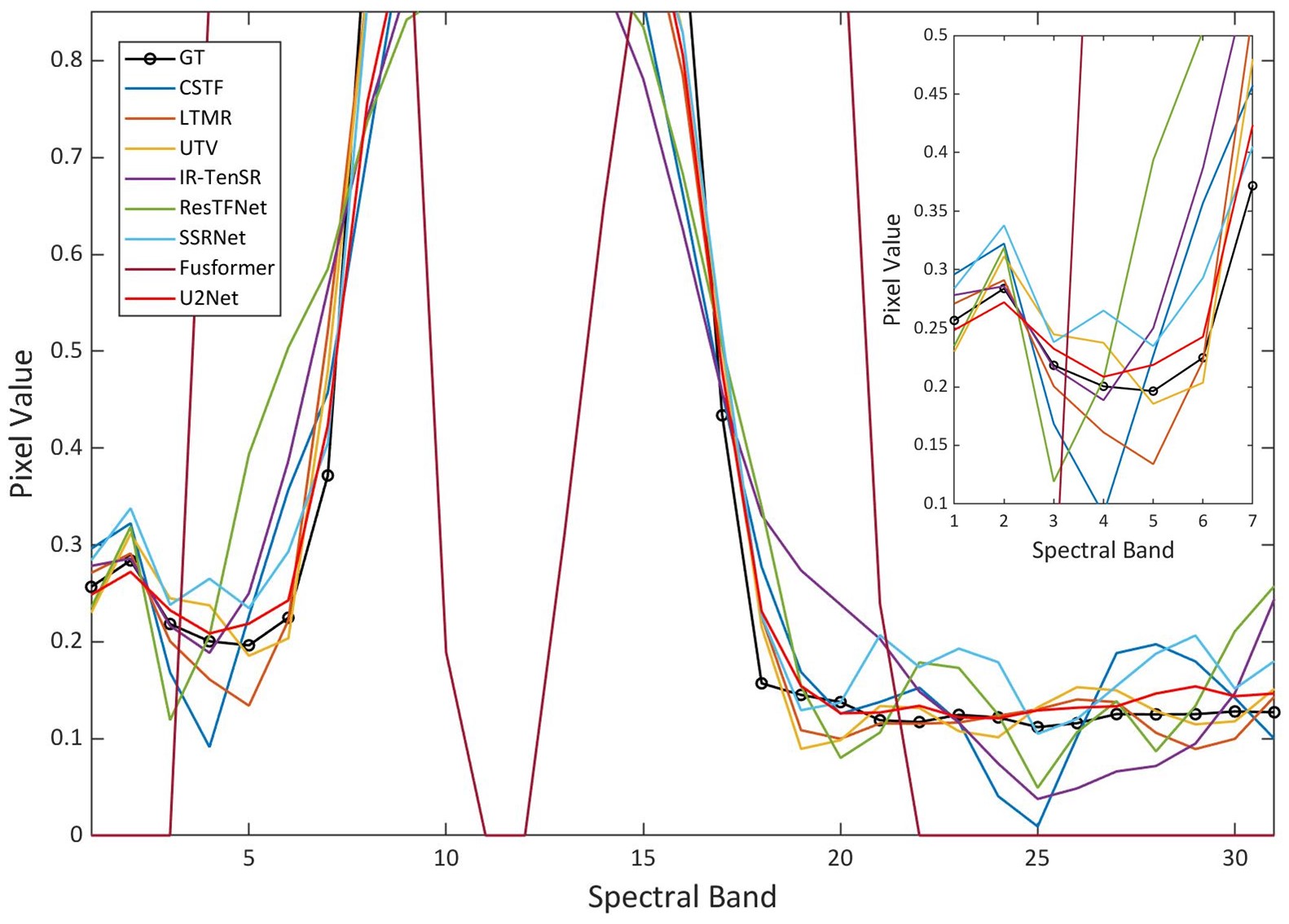}}
				{Spectral vectors at (286, 403)}
				\centering
				
			\end{minipage}
		\end{minipage}
	\end{center}
	\caption{The comparisons of spectral vectors from three spatial locations of a CAVE testing sample.\label{comparison6}}
\end{figure*}

\section{Experiments for HISR}
\label{s5}
\subsection{Experiment Settings}
\noindent\textbf{Datasets.} For the HISR task, experiments are conducted on the CAVE dataset \cite{2010Generalized}, which contains 31 RGB/LRHS image pairs with sizes $512\times512\times3$ and $512\times512\times31$. We select 20 samples for training, and the rest are for testing. The 20 training samples are made into 3920 overlapped RGB/LRHS/GT image pairs (80\% for training and 20\% for validation) with sizes $64\times64\times3$, $16\times16\times31$, and $64\times64\times31$, while the testing samples are processed as 11 RGB/LRHS/GT image pairs with sizes $512\times512\times3$, $128\times128\times31$, and $512\times512\times31$.

\noindent\textbf{Benchmarks and Evaluation Metrics.} We compare our method with some recent SOTA approaches, including five traditional methods: CSTF \cite{8359412}, LTMR \cite{dian2019hyperspectral}, LTTR \cite{2019learning}, UTV \cite{xujstars2020}, and IR-TenSR \cite{xu2022iterative}; and three DL-based works: ResTFNet \cite{2018Remote}, SSRNet \cite{9186332}, and Fusformer \cite{9841513}. Four commonly used metrics are selected, including PSNR, SSIM \cite{1284395}, SAM, and ERGAS \cite{2002Data}.

\noindent\textbf{Parameters Tuning.} 
For the HISR task, we set the values of $S$ and $S'$ in our network to 64 and 16, respectively. Upon training the U2Net, the initial learning rate, epoch, and batch size are set to 0.0003, 500, and 8, respectively. Additionally, we choose Adam as the optimizer, and the learning rate is halved every 50 epochs.

\begin{table}[t]	
	\centering\renewcommand\arraystretch{1.4}\setlength{\tabcolsep}{5.5pt}
	\footnotesize
	\caption{Quantitative evaluation results on 11 testing samples of the CAVE dataset. (\textcolor{red}{Red}: best; \textcolor{blue}{Blue}: second best).}	
	\begin{tabular}{cccccc}
		\toprule
		\textbf{Method} & PSNR($\pm$std) & SSIM($\pm$std) & SAM($\pm$std) & ERGAS($\pm$std) \\ 
		\midrule
		\textbf{CSTF} & 34.463$\pm$4.281 & 0.866$\pm$0.075 & 14.368$\pm$5.302 & 8.289$\pm$5.285 \\ 
		\textbf{LTMR} & 36.543$\pm$3.300 & 0.963$\pm$0.021 & 6.711$\pm$2.193 & 5.387$\pm$2.529 \\ 
		\textbf{LTTR} & 35.851$\pm$3.488 & 0.956$\pm$0.029 & 6.990$\pm$2.554 & 5.990$\pm$2.921 \\
		\textbf{UTV} & 38.615$\pm$4.064 & 0.941$\pm$0.043 & 8.649$\pm$3.376 & 4.519$\pm$2.817 \\
		\textbf{IR-TenSR} & 35.608$\pm$3.446 & 0.945$\pm$0.027 & 12.295$\pm$4.683 & 5.897$\pm$3.046\\
		\textbf{ResTFNet} & 45.584$\pm$5.465 & 0.994$\pm$0.006 & 2.764$\pm$0.699 & 2.313$\pm$2.438 \\ 
		\textbf{SSRNet} & 48.620$\pm$3.918 & \textcolor{blue}{0.995}$\pm$0.002& 2.542$\pm$0.837 & \textcolor{blue}{1.636}$\pm$1.219 \\ 
		\textbf{Fusformer} & \textcolor{blue}{49.983}$\pm$8.097 & 0.994$\pm$0.011 & \textcolor{blue}{2.203}$\pm$0.851 & 2.534$\pm$5.305 \\ 
		\textbf{U2Net} & \textbf{\textcolor{red}{50.441}}$\pm$4.403& \textbf{\textcolor{red}{0.997}}$\pm$0.002 & \textbf{\textcolor{red}{2.164}}$\pm$0.609 & \textbf{\textcolor{red}{1.267}}$\pm$0.967 \\ 
		\midrule
		\textbf{Ideal value} & \textbf{+$\infty$} & \textbf{1} & \textbf{0} & \textbf{0} \\ 
		\bottomrule
	\end{tabular}
	\label{cave} 
\end{table}

\subsection{Results on the Cave Dataset}
We assess the performance of recent SOTA approaches and our method on 11 testing samples of the CAVE dataset. The quantitative evaluation outcomes are presented in Tab.~\ref{cave}, and the U2Net achieves the best average results on all quality indicators. Additionally, the qualitative evaluation outcomes are shown in Fig.~\ref{comparison2} together with the GT. Obviously, our method exhibits the darkest AEM, proving its superiority in the HISR task. Furthermore, in Fig.~\ref{comparison6}, we display the spectral vectors from three different spatial locations of a testing sample. The spectral vectors of U2Net are the closest to the GT, indicating that our method has a potent spectral preservation ability.

\section{Conclusion}
\label{s6}
In this paper, we propose a spatial-spectral-integrated double U-shape network called U2Net for image fusion tasks. The U2Net employs a spatial U-Net and a spectral U-Net to extract spatial details and spectral characteristics discriminately and hierarchically. Besides, we create a novel structure named S2Block that sufficiently merges feature maps from diverse images in a logical and comprehensive manner. We compare our U2Net with several recent SOTA pansharpening and HISR approaches. The proposed method outperforms all others on a series of datasets, demonstrating its exceptional feature learning, information integration, and generalization capabilities. Therefore, we are confident that our method offers an effective solution for the image fusion problems.

\begin{acks}
This research is supported by NSFC (12271083), Natural Science Foundation of Sichuan Province (2022NSFSC0501).
\end{acks}

\bibliographystyle{ACM-Reference-Format}
\balance
\bibliography{sample-base}

\appendix

\end{document}